\documentclass[preprint,12pt]{elsarticle}

\pdfoutput=1
\usepackage{amssymb}
\usepackage{amsmath}
\usepackage{mathrsfs}
\usepackage{url}
\usepackage[colorlinks,bookmarksopen,bookmarksnumbered,citecolor=steelblue,urlcolor=red]{hyperref}
\usepackage{algorithm}
\usepackage[noend]{algpseudocode}
\usepackage{xargs}
\usepackage{todonotes}
\usepackage{amsthm}
\usepackage{multirow}
\usepackage{caption, subcaption}
\usepackage{tikz}
\usetikzlibrary{external}
\usetikzlibrary{shapes,arrows,calc,angles,quotes,arrows.meta}
\usepackage{pgfplots}
\usepgfplotslibrary{fillbetween}
\pgfplotsset{compat=1.17}
\usetikzlibrary{external}
\newtheorem{definition}{Definition}

\definecolor{steelblue}{RGB}{70,130,180}
\renewcommand{\algorithmiccomment}[1]{\State\bgroup//~#1\egroup}
\algnewcommand\algorithmicforeach{\textbf{for each}}
\algdef{S}[FOR]{ForEach}[1]{\algorithmicforeach\ #1\ \algorithmicdo}
\newcommandx{\iman}[2][1=]{\todo[linecolor=blue,backgroundcolor=blue!25,bordercolor=blue,size=small,author=Iman,#1]{#2}}
\newcommandx{\venki}[2][1=]{\todo[linecolor=red,backgroundcolor=red!25,bordercolor=red,size=small,author=Venki,#1]{#2}}
\newcommandx{\sleiman}[2][1=]{\todo[linecolor=green,backgroundcolor=green!25,bordercolor=green,size=small,author=Sleiman,#1]{#2}}

\def\bbr{\mathbb R}

\def\bbs{\mathbb S}
\def\bbp{\mathbb P}
\def\bbe{\mathbb E}

\def\calx{\mathcal X}

\DeclareMathOperator*{\argmin}{argmin}

\newcommand{\rrtstar}{\texttt{RRT}^{\star}}
\newcommand{\nrbrrtstar}{\texttt{NRB-RRT}^{\star}}

\newcommand{\rlpar}[1]{\left ( {#1} \right)}

\journal{Artificial Intelligence}

\begin{document}

\begin{frontmatter}



\title{Risk Bounded Nonlinear Robot Motion Planning With Integrated Perception \& Control}

\tnotetext[t1]{This work was completed when V. Renganathan was a PhD student at The University of Texas at Dallas, USA.}

\affiliation[inst1]{
organization={Department of Automatic Control, Lund University},
addressline={Naturvetarvägen 18}, 
city={Lund},
postcode={221 00}, 
state={SE},
country={Sweden.}}

\affiliation[inst2]{
organization={Eric Jonsson School of Engineering \& Computer Science, The University of Texas at Dallas},
addressline={800 W Campbell Rd}, 
city={Richardson},
postcode={75080}, 
state={TX},
country={USA.}}

\affiliation[inst3]{organization={School of Engineering},
addressline={Massachusetts Institute of Technology}, 
city={Cambridge},
postcode={02139}, 
state={MA},
country={USA.}}

\affiliation[inst4]{organization={CIICADA Lab, School of Engineering},
addressline={The Australian National University}, 
city={Acton},
postcode={0200}, 
state={ACT},
country={Australia.}}

\author[inst1]{Venkatraman Renganathan}
\ead{venkat@control.lth.se}

\author[inst2]{Sleiman Safaoui}
\ead{sleiman.safaoui@utdallas.edu}

\author[inst3]{Aadi Kothari}
\ead{aadi@mit.edu}

\author[inst2]{Benjamin Gravell}
\ead{benjamin.gravell@utdallas.edu}

\author[inst4]{Iman Shames}
\ead{iman.shames@anu.edu.au}

\author[inst2]{Tyler Summers}
\ead{tyler.summers@utdallas.edu}

\begin{abstract}
Robust autonomy stacks require tight integration of perception, motion planning, and control layers, but these layers often inadequately incorporate inherent perception and prediction uncertainties, either ignoring them altogether or making questionable assumptions of Gaussianity. Robots with nonlinear dynamics and complex sensing modalities operating in an uncertain environment demand more careful consideration of how uncertainties propagate across stack layers. We propose a framework to integrate perception, motion planning, and control by explicitly incorporating perception and prediction uncertainties into planning so that risks of constraint violation can be mitigated. Specifically, we use a nonlinear model predictive control based steering law coupled with a decorrelation scheme based Unscented Kalman Filter for state and environment estimation to propagate the robot state and environment uncertainties. Subsequently, we use distributionally robust risk constraints to limit the risk in the presence of these uncertainties. Finally, we present a layered autonomy stack consisting of a nonlinear steering-based distributionally robust motion planning module and a reference trajectory tracking module. Our numerical experiments with nonlinear robot models and an urban driving simulator show the effectiveness of our proposed approaches. 
\end{abstract}




\end{frontmatter}

\section*{Supplementary Material} \label{sec_supplements}
Video of the simulation is available at \url{https://youtu.be/KpyWXRZ-wSI} and the simulation code is available at \url{https://github.com/TSummersLab/Risk_Bounded_Nonlinear_Robot_Motion_Planning}.

\section*{Acknowledgement}
This work is partially supported by Defence Science and Technology Group, through agreement MyIP: ID10266 entitled \textbf{``Hierarchical Verification of Autonomy Architectures''}, the Australian Government, via grant AUSMURIB000001 associated with ONR MURI grant N00014-19-1-2571, and by the United States Air Force Office of Scientific Research under award number FA2386-19-1-4073.

\section{Introduction}
Safety is a critical issue for robotic and autonomous systems that must traverse through uncertain environments. More sophisticated motion planning and control algorithms are needed as environments become increasingly dynamic and uncertain to ensure safe and effective autonomous behavior. Safely deploying robots in such dynamic environments requires a systematic accounting of various risks both within and across layers in an autonomy stack from perception to motion planning and control. Many widely used motion planning algorithms have been developed in deterministic settings. However, since motion planning algorithms must be coupled with the outputs of inherently uncertain perception systems, there is a crucial need for more tightly coupled perception and planning frameworks that explicitly incorporate perception and prediction uncertainties.

Motion planning under uncertainty has been considered in several lines of recent research \cite{blackmore_pioneer,agha_firm, luders_rrt, luders_rrtstar, blackmore_cc_mp, liu_risk_aware_mp,zhu2019chance}. Many approaches make questionable assumptions of Gaussianity and utilize chance constraints, ostensibly to maintain computational tractability. However, this can cause significant miscalculations of risk, and the underlying risk metrics do not necessarily possess desirable coherence properties \cite{rockafellar2007coherent, pavone_risk}. The emerging area of distributionally robust optimization (DRO) shows that stochastic uncertainty can be handled in much more sophisticated ways without necessarily sacrificing computational tractability \cite{dr_goh}. These approaches allow modelers to explicitly incorporate inherent ambiguity in probability distributions, rather than making overly strong structural assumptions on the distribution. 

Traditionally, the perception and planning components in a robot autonomy stack are loosely coupled in the sense that nominal estimates from the perception system may be used for planning, while inherent perception uncertainties are usually ignored. This paradigm is inherited, in part, from the classical separation of estimation and control in linear systems theory. However, in the presence of uncertainties and constraints, estimation and control should \emph{not} be separated as there are needs and opportunities to explicitly incorporate perception uncertainties into planning, both to mitigate risks of constraint violation \cite{blackmore_pioneer,florenceIPC,luders_rrt,summers_iros_2018,zhu2019chance} and to actively plan paths that improve perception \cite{davide_perception_plan}. 

The present paper is aimed towards establishing a systematic framework for integrating the perception and control components using a coherent risk assessment with distributionally robust risk constraints. Algorithms for motion planning under uncertainty have begun to address mission safety, state and control constraints, and trajectory robustness using chance constraints, as in \cite{luders_rrt}. This approach has been recently broadened in \cite{summers_iros_2018, renganathan2020integrated, hakobyan2020wasserstein} using a more general framework of axiomatic risk theory and distributionally robust optimization. 

\noindent \textit{Contributions:} This manuscript is a significant extension of our previous works \cite{renganathan2020integrated, safaoui2021risk}. In this paper, we relax the assumptions from our previous works by considering both motion model and sensor models to be nonlinear with additive uncertainties and propose an unified framework aimed toward a tighter integration of perception and planning in autonomous robotic systems. Our main contributions are: 
\begin{itemize}
    \item We propose a distributionally robust incremental sampling-based motion planning framework that explicitly and coherently incorporates perception and prediction uncertainties. Our solution approach called Nonlinear Risk Bounded $\rrtstar$ $(\nrbrrtstar)$ (Algorithm 1), approximates asymptotically optimal risk-bounded trajectories. 
    \item We design output feedback policies and consider moment-based ambiguity sets of distributions to enforce probabilistic collision avoidance constraints under the worst-case distribution in the ambiguity set (Algorithm 2). Our output feedback policy satisfying input and state risk constraints is constructed using a nonlinear model predictive controller coupled with an Unscented Kalman Filter for state estimation.
    \item We formulate distributionally robust collision avoidance risk constraints. We demonstrate via numerical simulation results that this gives a more sophisticated and coherent risk quantification compared to an approach that accounts for uncertainty using Gaussian assumptions, without increasing the computation complexity. 
    \item We demonstrate our proposed algorithms and approaches using a unicycle model and a bicycle model in an open urban driving simulator \cite{carla_sim} and show that risk bounded motion planning can be achieved effectively for nonlinear robotic systems.
\end{itemize} 
The rest of the paper is organized as follows. The integrated environment state formulation using the nonlinear dynamical model of the robot and the obstacle and their uncertainty modeling is discussed in \S\ref{sec_robot_model}. Next, we describe the layers of a distributionally robust autonomy stack and formulate a general stochastic optimal control problem in \S\ref{sec_dr_mp}. An output feedback based steering law with uncertainty propagation is presented in \S\ref{sec_steer}. In \S\ref{sec_drrrt_algorithm}, we describe the planner module that uses the $(\nrbrrtstar)$ motion planning algorithm for planning a reference trajectory. Then, in \S\ref{sec_dr_collision_check}, we discuss about the reformulation technique for the distributionally robust risk constraints that define obstacle avoidance. The simulation results are then presented in \S\ref{sec_sim_results}. Finally, the paper summary and future research directions are discussed in \S\ref{sec_conclusions}.

\section*{Notation}
The set of real numbers and natural numbers are denoted by $\mathbb{R}$ and $\mathbb{N}$, respectively. The subset of natural numbers between and including $a$ and $b$ with $a < b$ is denoted by $[a:b]$. The operators $|\cdot|$ and $(\cdot)^c$ denote the set cardinality and the set complement of its argument, respectively. The operators $\oplus, \backslash$ denote the set translation and set subtraction respectively. An identity matrix in dimension $n$ is denoted by $I_{n}$. For a non-zero vector $x \in \bbr^{n}$ and a matrix $P \in \bbs^{n}_{++}$ (set of positive definite matrices), let $\left \| x \right \Vert_{P} = \sqrt{x^{\top} P x}$. 

\section{Description of the Robot \& the Environment} \label{sec_robot_model}
Consider a robot operating in an environment, $\mathcal{X} \subset \bbr^{n}$ cluttered with obstacles whose locations and motion are uncertain. The robot is assumed to be equipped with sensors and a perception system that support safe navigation through the environment. In what follows, we present the necessary ingredients for presenting the results such as the models for the robot, the environment, and the perception system.
\subsection{Robot \& Environment Model} 
\subsubsection{Robot Model:}
The robot is modeled as a stochastic discrete-time dynamical system:
\begin{align} \label{eqn_robot_dynamics}
    x_{t + 1} &= \mathbf{f}(x_t, u_t) +  w_{r,t}, 
\end{align}
where $x_t \in \bbr^n, u_t \in \bbr^m$ are the robot state and input respectively at time $t$ and $\mathbf{f}: \bbr^{n} \times \bbr^{m} \rightarrow \bbr^{n}$ is a nonlinear function that represents the robot dynamics. The initial condition $x_0$ is subject to an uncertainty model with the true distribution $\bbp_{x_0}$ of $x_0$ belonging to a moment-based ambiguity set $\mathcal{P}^{x_{0}}$, i.e., $\bbp_{x_0} \in \mathcal{P}^{x_{0}}$ where
\begin{align} \label{eqn_ambig_x0}
\mathcal{P}^{x_{0}} := \left\{ \bbp_{x_0} \mid \bbe[x_{0}] = \bar{x}_{0}, \bbe[(x_{0} - \bar{x}_{0}) (x_{0} - \bar{x}_{0})^{\top}] = \Sigma_{x_{0}}  \right\},    
\end{align}
and $\bar{x}_0$ and $\Sigma_{x_{0}}$ are some known parameters of appropriate dimensions.
The additive process noise $w_{r,t} \in \bbr^n$ is a zero-mean random vector independent and identically distributed across time according to some prescribed distribution $\bbp_{w_{r}}$ with covariance $\Sigma_{w_{r}}$.
\subsubsection{Environment Model}
The environment $\mathcal{X}$ is assumed to be convex and represented by finite number of linear inequalities
\begin{align} \label{eqn_environment_represent}
    \mathcal{X} := \left\{ x_t \mid A_e x_t \leq b_e \right\} \subset \bbr^{n},
\end{align}
where the matrices $A_e, b_e$ are of appropriate dimensions. We collectively refer the set of obstacles in the environment to be avoided as $\mathcal{B}$ with $|\mathcal{B}| = F \in \mathbb{N}$. We represent the shape of an obstacle $i \in \mathcal{B}$ at a given initial time $(t_0 = 0)$ by $\mathcal{O}_{i,0} \subset \bbr^{n}$. For simplicity, $\forall i \in \mathcal{B}$ the obstacles shapes $\mathcal{O}_{i,0}$ are assumed to be convex polytopes. That is, for $i \in \mathcal{B}$
\begin{align} \label{eqn_obs_linear_inequalities}
    \mathcal{O}_{i,0} := \left\{ x_{0} \mid A_i x_{0} \leq b_{i, 0} \right\} \subset \bbr^{n},
\end{align}
with the matrices $A_i \in \bbr^{n \times n}, b_{i,t} \in \bbr^{n}$. All obstacles are assumed to have an unpredictable motion and the set defining the space occupied by the obstacle $i \in \mathcal{B}$ at any time $t$ is given by $\mathcal{O}_{i,t}$. 
Then, the free space in the environment at time $t$ is
\begin{align}
    \mathcal{X}^{\texttt{free}}_{t} &:= \mathcal{X} \, \big\backslash \, \bigcup_{i \in \mathcal{B}} \mathcal{O}_{i,t}, \quad \text{where,} \\
    \mathcal{O}_{i,t} &= \mathsf{R}_{i,t-1} \mathcal{O}_{i,t-1} \oplus \{ \bar{r}_{i,t} \} \oplus \{ r_{i,t-1} \}, \quad \forall t \geq 1. \label{eqn_obstacle_set_description}
\end{align}
Here, $\bar{r}_{i,t} \in \bbr^{n}$ represents a known nominal translation and $r_{i,t} \in \bbr^{n}$ is a zero-mean random vector that represents the uncertain translation (possibly an unknown location or unpredictable obstacle motion) of the obstacle $i \in \mathcal{B}$ at time $t$, and it is assumed to follow a known distribution $\bbp^{r}_{i,t}$ with covariance $\Sigma^{r}_{it}$. Further, $\mathsf{R}_{i,t-1} \in \bbr^{n \times n}$ denotes the product of random rotation matrices that represents the uncertain rotation of the obstacle $i \in \mathcal{B}$ at time $t$ where we used the definition that multiplying a matrix $\mathsf{R}_{i,t}$ to a set $\mathcal{O}_{i,t}$ is defined by $\mathsf{R}_{i,t} \mathcal{O}_{i,t} := \left\{ \mathsf{R}_{i,t} \cdot (o - c_{o,i}) + c_{o,i} \mid o \in \mathcal{O}_{i,t} \right\}$ with $c_{o,i} \in \bbr^{n}$ denoting the centroid of obstacle set $\mathcal{O}_{i,t}$. Let us denote the vectorized version of the uncertain rotation matrix $\mathsf{R}_{i,t}$ as $\mathsf{r}_{i,t} := \mathsf{vec}\left(\mathsf{R}_{i,t}\right) \in \bbr^{n^{2}}$ and it is assumed to be a zero-mean random vector following a known distribution $\bbp^{\mathsf{r}}_{i,t}$ with covariance $\Sigma^{\mathsf{r}}_{it}$. However, for the ease of exposition in the remainder of this paper, we shall assume that either $\mathsf{R}_{i,t}$ is known or $\mathsf{R}_{i,t}$ being an identity transformation (no rotation) at all time steps $t$, so that there is uncertainty only in the translation. We denote by $\mathsf{X}_{i,t} \in \bbr^{l},$ the states of the obstacle $i \in \mathcal{B}$, which for instance could represent the position and velocity of the centroid of the obstacle at time $t$. Then, the evolution of the state of the obstacle $i \in \mathcal{B}$ can be written as 
\begin{align}
    \mathsf{X}_{i,t+1} 
    &= \mathsf{g}_{i, t} \left( \mathsf{X}_{i,t} \right) + w_{\mathcal{O}, i, t}. 
\end{align}
The random vector $r_{i,t}$ that resulted in the set translation in \eqref{eqn_obstacle_set_description} manifests itself as the zero-mean process noise that affects the states of the obstacle $i \in \mathcal{B}$ as $w_{\mathcal{O}, i, t} \in \bbr^{l}$ and $w_{\mathcal{O}, i, t}$ is assumed to follow a known distribution $\bbp_{w_{\mathcal{O}_{i}}}$ with covariance $\Sigma_{w_{\mathcal{O}_{i}}}$. Further, $\mathsf{g}_{i,t}: \bbr^{l} \rightarrow \bbr^{l}$ denotes the function (possibly nonlinear) that represents the dynamics of the obstacle state at time $t$. For instance, the obstacle can be assumed to travel at a constant velocity but with process noise. Further, we assume that the uncertainty in the motion of obstacle $i$ is independent from every other obstacle $j \in \mathcal{B}$, which implies $\bbe\left[ w_{\mathcal{O}, i, t} w^{\top}_{\mathcal{O}, j, t} \right] = 0, \forall i, j \in \mathcal{B}, i \neq j$. Then, the joint evolution of all obstacles can be written as 
\begin{align}
    \mathsf{X}_{\mathcal{O},t+1} = \mathbf{g}\left( \mathsf{X}_{\mathcal{O},t} \right) +  w_{\mathcal{O}, t}, 
\end{align}
where $\mathsf{X}_{\mathcal{O}_{t}}$, $\mathbf{g}(\cdot)$, and  $w_{\mathcal{O}, t}$ represent the concatenated states, nonlinear dynamics and the process noises of all the $F$ obstacles at time $t$ respectively. The additive process noise $w_{\mathcal{O}, t} \in \bbr^{Fl}$ is a zero-mean random vector independent and identically distributed across time according to some prescribed distribution $\bbp_{w_{\mathcal{O}}}$ with covariance $\Sigma_{w_{\mathcal{O}}}$. We assume that the process noises affecting the robot and the obstacles $ i \in \mathcal{B}$ are independent of each other. That is, $\bbe\left[w_{r,t} w^{\top}_{\mathcal{O}, i, t} \right] = 0, \forall i \in \mathcal{B}$.

\subsubsection{Integrated Environmental Dynamics} 
We concatenate both the robot's state and the obstacle states at time $t$ to form the environmental state
\begin{align} \label{eqn_environment_state}
    \mathcal{Z}_t &= \begin{bmatrix} x_t \\  \mathsf{X}_{\mathcal{O}_{t}} \end{bmatrix} \in \bbr^{n_z}, \quad n_z = n+Fl.
\end{align}
Then the dynamics of the environmental state is given by
\begin{align}\label{eqn_env_st_dynamics}
    \mathcal{Z}_{t+1} &= \underbrace{\begin{bmatrix}\mathbf{f}(x_t, u_t) \\ \mathbf{g}(\mathsf{X}_{\mathcal{O}_{t}}) \end{bmatrix}}_{\tilde{f}\left(\mathcal{Z}_{t}, u_t \right)} +  \underbrace{\begin{bmatrix} w_{r,t} \\ w_{\mathcal{O}, t} \end{bmatrix}}_{w_t}.
\end{align}
The initial condition $\mathcal{Z}_0$ is subject to an uncertainty model with the true distribution $\bbp_{\mathcal{Z}_0}$ of $\mathcal{Z}_0$ belonging to a moment-based ambiguity set $\mathcal{P}^{\mathcal{Z}_{0}}$, i.e., $\bbp_{\mathcal{Z}_0} \in \mathcal{P}^{\mathcal{Z}_{0}}$ where
\begin{align} \label{eqn_ambig_Z0}
\mathcal{P}^{\mathcal{Z}_{0}} := \left\{ \bbp_{\mathcal{Z}_0} \mid \bbe[\mathcal{Z}_{0}] = \bar{\mathcal{Z}}_{0}, \bbe[(\mathcal{Z}_{0} - \bar{\mathcal{Z}}_{0}) (\mathcal{Z}_{0} - \bar{\mathcal{Z}}_{0})^{\top}] = \Sigma_{\mathcal{Z}_{0}}  \right\}.    
\end{align}
Here, $w_{t} \in \bbr^{n_{z}}$ is a zero-mean random vector independent and identically distributed across time with covariance $\Sigma_{w} = \begin{bmatrix} \Sigma_{w_{r}} & 0 \\ 0 & \Sigma_{w_{\mathcal{O}}} \end{bmatrix}$. At any instant of time $t$, $x_t$ can be extracted from the environmental state $\mathcal{Z}_{t}$ as
\begin{align} \label{eqn_x=CZ}
    x_t &= \underbrace{\begin{bmatrix} I_{n} & \mathbf{0}_{n \times Fl} \end{bmatrix}}_{C_{xr}} \mathcal{Z}_{t}.
\end{align}

\subsection{Sensor Model}\label{sec:sensor}
In an autonomous robot, the environmental state $\mathcal{Z}_t$ must be estimated with a  perception system represented by a noisy on-board sensor measurements. We assume that a high-level perception system, such as Semantic SLAM described in \cite{semantic_slam} is in place and it processes high dimensional raw data $\Theta_{t} \in \bbr^{N_{\Theta}}, N_{\Theta} \in \mathbb{N}$ to recognize the robot and the obstacles to produce noisy joint measurements of their respective states. In particular, the perception system recognizes the robot and obstacles through their distinctive features through mappings $\Upsilon_{x}: \bbr^{N_{\Theta}} \rightarrow \bbr^{n}, \Upsilon_{\mathcal{O}}: \bbr^{N_{\Theta}} \rightarrow \bbr^{l}$, labels the entities accordingly and returns a noisy measurement about their position and orientation. We abstract this whole process and assume that the perception system produces noisy measurements of the robot and obstacle states using an assumed nonlinear output model
\begin{align} \label{eqn_output_sensor_model}
y_t &= \mathcal{S}(\mathcal{Z}_t) + v_t, 
\end{align}
where $y_t \in \bbr^{p}$ is the output measurement and $\mathcal{S}: \bbr^{n_{z}} \rightarrow \bbr^{p}$ is a nonlinear function. The additive measurement noise $v_{t} \in \bbr^p$ is a zero-mean random vector independent and identically distributed across time according to some prescribed distribution $\bbp_{v}$ with covariance $\Sigma_{v}$. \\

There can be interactions between the uncertainties of the perception system and the process disturbance, due to coupling between robot and environmental states in the perception and the use of an output feedback control policy. Often, this interaction is ignored for the sake of simplicity and as a result can impose undue risk. To model the interaction, we assume that the process noise $w_t$ and the measurement noise $v_{t}$ are cross correlated with each other. That is, the joint noise $\begin{bmatrix} w_{t} \\ v_{t} \end{bmatrix}$ has zero mean with covariance
\begin{align} \label{eqn_w_v_cross_correl}
    \Sigma_{wv} = \bbe\left[\begin{bmatrix} w_{t} \\ v_{t} \end{bmatrix} \begin{bmatrix} w_{t} \\ v_{t} \end{bmatrix}^{\top}\right] = \begin{bmatrix} \Sigma_{w} & M \\ M^{\top} & \Sigma_{v} \end{bmatrix} \succ 0,
\end{align}
where $M = \bbe[w_t v^{\top}_t] \in \bbr^{n_{z} \times p}$ denotes the rank-one cross-correlation matrix across all time steps $t$ and the matrix $M \neq \mathbf{0}_{n_{z} \times p}$ is such that the joint noise covariance $\Sigma_{wv}$ given by \eqref{eqn_w_v_cross_correl} is positive definite.
\subsection{State \& Input Constraints}
The robot is nominally subject to constraints on the state and input of the form, $\forall t = 0,\dots,T-1$,
\begin{align}
x_t &\in \mathcal{X}^{\texttt{free}}_{t}, \label{eqn_constraints_x}\\
u_t &\in \mathcal{U}, \label{eqn_constraints_u}
\end{align}
where the $\mathcal{U} \subset \bbr^{m}$ are assumed to be convex polytope. We can rewrite the set description of the required sets 
using the robot inputs and environmental state as
\begin{align}
    \mathcal{U} &= \{ u_t \mid A_{u} u_t \leq b_{u}\}, \label{eqn_u_cons}\\
    \mathcal{X} &= \{ \mathcal{Z}_{t} \mid A_{0} C_{xr} \mathcal{Z}_{t} \leq b_{0}\}, \\
    \mathcal{O}_{it} &= \{ \mathcal{Z}_{t} \mid A_{i} C_{xr} \mathcal{Z}_{t} \leq b_{it}\}, \quad i \in \mathcal{B}
\end{align}
where $b_u \in \bbr^{n_u}, b_0 \in \bbr^{n_E}, b_{it} \in \bbr^{n_i},$ and $A_u, A_0,$ and $A_i$ are matrices of appropriate dimension. Suppose that $n_{ob_i}$ and $n_{env}$ be the number of constraints for obstacle $i \in \mathcal{B}$ and the environment $\calx$ respectively. Then, the total number of constraints is denoted by
\begin{align}
 n_{total} = n_{env} + \sum_{i=1}^{F}n_{ob_i}.   
\end{align}

\section{Layers of a Distributionally Robust Autonomy Stack} \label{sec_dr_mp}
\begin{figure}[H]
    \centering
    \includegraphics[scale=0.38]{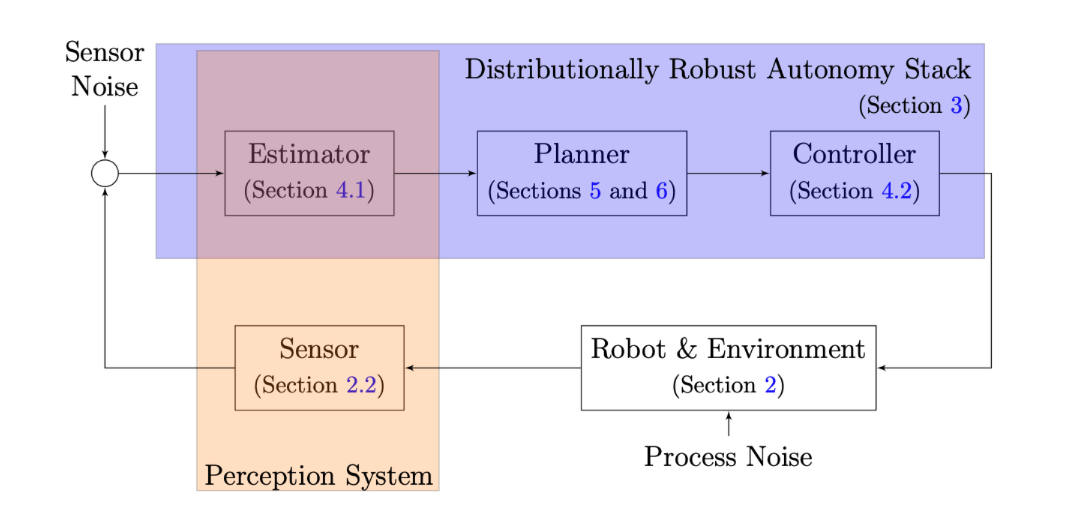}
\caption{A distributionally robust autonomy stack with integrated perception, planning, and control layers.}
    \label{fig_dr_autonomy_stack}
\end{figure}

The autonomy stack in an autonomous robot, as depicted in Figure~\ref{fig_dr_autonomy_stack}, can be partitioned into a hierarchy of i) a perception system that jointly estimates the robot and environmental states; ii) a motion planner which generates a reference trajectory through the environment, and iii) a feedback controller that tracks the reference trajectory online to mitigate the effects of disturbances. To emphasize our integrated approach, we first pose a general stochastic optimal control problem that incorporates all of these layers. Then we describe how the layers are integrated with explicit incorporation of uncertainties across layers in subsequent sections.

Given an initial state distribution $\mathcal{Z}_0 \sim \bbp_{\mathcal{Z}_0}$ and a set of final goal locations $\mathcal{X}_{goal} \subset \calx$, we find a measurable output-and-input-history-dependent control policy $\pi = [\pi_{0}, \dots \pi_{T-1}]$ with $u_{t} = \pi_{t}(y_{[0:t]}, u_{[0:t-1]})$ that moves the robot state mean to the goal set while respecting input constraints and distributionally robust collision risk constraints, while incorporating the ambiguities in the distributions described above. To this aim, we define the following distributionally robust constrained stochastic optimal control problem 
\begin{subequations} \label{eqn_dr_motion_planning_new}
    \begin{align}
        &\underset{\pi}{\text{minimize}} & & \left[\sum^{T-1}_{t = 0} \ell_t(\bbe[\mathcal{Z}_t], u_t) + \ell_T(\bbe[\mathcal{Z}_T])\right] \label{subeqn_cost_fn} \\
        &\text{subject to} & &  \mathcal{Z}_{t + 1} = \tilde{f}(\mathcal{Z}_{t}, u_t) + w_t, \\
        & & & y_{t} = \mathcal{S}(\mathcal{Z}_t) + v_t, \\
        & & & \mathcal{Z}_0 \sim \bbp_{\mathcal{Z}_0}\in \mathcal{P}^\mathcal{Z}, \\
        & & & w_t \sim \bbp_{w}(0, \Sigma_{w}), v_t \sim \bbp_{v}(0, \Sigma_{v}), \\
        & & & u_t \in \mathcal{U} = \{ u_t \mid A_{u} u_t \leq b_{u}\}, \\
        & & & \underset{\bbp_{\mathcal{Z}_t}\in \mathcal{P}^{\mathcal{Z}}}{\sup} \mathcal{R} \left( C_{xr} \mathcal{Z}_{t} \notin \mathcal{X}^{\texttt{free}}_{t} \right) \leq \alpha_{t}, \quad \forall t \in [0:T], \label{subeqn_risk_metric}
    \end{align}
\end{subequations}
where $\mathcal{P}^\mathcal{Z}$ is an ambiguity set of marginal state distributions, $\alpha_{t} \in (0,0.5]$ is a user-prescribed stage risk parameter, and $\mathcal{R}(\cdot)$ is a risk measure that operates on the distribution of states and satisfies certain desirable axioms mentioned in \cite{pavone_risk}. The stage cost functions $\ell_t(\cdot)$ penalizes the robot’s distance to the goal set and actuator effort, and are assumed to be expressed in terms of the environmental state mean $\bbe[\mathcal{Z}_t]$, so that the state uncertainty appears only in the constraints. 

Two key features distinguish our problem formulation. First, the state constraints are expressed as \emph{distributionally robust risk constraints}, using the risk measure $\mathcal{R}(\cdot)$ and the ambiguity set $\mathcal{P}^\mathcal{Z}$ whose construction will be detailed below. Since we will work with moment based ambiguity sets for the states, we give special consideration to the distributionally robust value-at-risk (DR-VaR) risk measure
\begin{align} \label{eqn_DR_Risk_Constraint}
    \underset{\bbp_{\mathcal{Z}_t}\in \mathcal{P}^{\mathcal{Z}}}{\sup} \bbp_{\mathcal{Z}_t} (C_{xr} \mathcal{Z}_{t} \notin \mathcal{X}^{\texttt{free}}_{t}) \leq \alpha_{t}, \quad \forall t \in [0:T],
\end{align}
which is a special case of \eqref{subeqn_risk_metric} with a tractable reformulation. This means that the nominal constraints $x_t \in \mathcal{X}^{\texttt{free}}_{t}$ are enforced with probability $\alpha_t$ at time $t$ under the worst-case distribution in the ambiguity set and thus making them \emph{distributionally robust chance constraints}. 

Second, since information about the environmental state is obtained only from noisy measurements, we optimize over dynamic output feedback policies. This policy encompasses the full autonomy stack from perception to planning to control. However, the problem is infinite-dimensional, and moreover, the distributionally robust risk constraint in \eqref{eqn_DR_Risk_Constraint} is also infinite-dimensional and inherently non-convex due to the underlying obstacle avoidance constraints. Thus, solving \eqref{eqn_dr_motion_planning_new} exactly is essentially impossible. Instead, we aim for an approximate solution. Our proposed solution framework, detailed below, integrates a dynamic nonlinear state estimator (the estimator block in Figure~\ref{fig_dr_autonomy_stack}), a nonlinear feedback controller (the controller block in Figure~\ref{fig_dr_autonomy_stack}), and a sampling-based kinodynamic motion planning under uncertainty algorithm (the planner block in Figure~\ref{fig_dr_autonomy_stack}). Specifically, in this work we propose using an Unscented Kalman filter as the nonlinear estimator, a nonlinear model predictive control as the nonlinear feedback controller, and a novel distributionally robust, kinodynamic variant of the $\rrtstar$ called Nonlinear Risk Bounded $\rrtstar (\nrbrrtstar)$ as the planner\footnote{It is worth noting that other choices for estimator, controller, and planner subsystems can be incorporated in our proposed framework, nevertheless, in this work, for reasons that will be explained later, we make the aforementioned choices for these subsystems.}. This integration and explicit incorporation of state estimation uncertainty into the motion planning and control takes a step toward tighter integration of perception, planning, and control, which are nearly always separated in state-of-the-art robotic systems. 

\section{Output Feedback Based Steering Law: Unscented Kalman Filter with Nonlinear Model Predictive Controller} \label{sec_steer}
Sampling based motion planning algorithms require a steering law to steer the robot from one pose to another feasible pose in the free space. Since the environment state has nonlinear dynamics and must be estimated from noisy nonlinear output measurements \eqref{eqn_output_sensor_model}, our proposed steering law $\pi = [\pi_0,\dots, \pi_{T-1}]$ with $u_t = \pi_t(y_{0:t}, u_{0:t-1})$ comprises a combination of nonlinear dynamic state estimator and a nonlinear feedback controller. We utilize the Unscented Kalman Filter (UKF) for joint estimation of robot and obstacle states. The UKF provides derivative-free estimation of the first and second moment, which can be more accurate than the Extended Kalman Filter while remaining computationally tractable. A nonlinear model predictive controller as described in \cite{gu2005}, \cite{Allgower_NMPC} is then utilized based on the state estimate from the UKF. 

\subsection{The Unscented Kalman Filter (UKF)}\label{sec:filter}
In this subsection, we elaborate about the Unscented Kalman Filter algorithm which serves as an estimator module for a perception system present in the autonomy stack shown in Figure \ref{fig_dr_autonomy_stack}. Later in subsection \ref{sec:NLMPC}, we will describe how the estimates from UKF are then used to realize an output feedback based controller. \\

\textbf{Noise Decorrelation.} Before describing the UKF, we present a decorrelation scheme \cite{ukf_correl_noises} to handle cross-correlation between process noise and sensor induced by the coupling between perception and control layers from \eqref{eqn_w_v_cross_correl}.  The scheme uses a pseudo state process equation to reconstruct a corresponding pseudo process noise, which is no longer correlated with the measurement noise.  It is evident from \eqref{eqn_output_sensor_model} that
\begin{align}
    y_t - \mathcal{S}(\mathcal{Z}_t) - v_t = 0.
\end{align}
Thus the environmental dynamics given in \eqref{eqn_env_st_dynamics} can be rearranged as follows
\begin{align}
    \mathcal{Z}_{t+1} &= \tilde{f}(\mathcal{Z}_{t}, u_t) + w_{t} + \mathcal{H} (y_t - \mathcal{S}(\mathcal{Z}_t) - v_t) \\
    &= \underbrace{\tilde{f}(\mathcal{Z}_{t}, u_t) - \mathcal{H} \mathcal{S}(\mathcal{Z}_t) + \mathcal{H} y_t}_{f^{*}(\mathcal{Z}_{t}, u_t)} + \underbrace{w_{t} - \mathcal{H} v_t}_{w^{*}_{t}}, \label{eqn_pseudo_env_dyn}
\end{align}
where $f^{*}(\mathcal{Z}_{t}, u_t), w^{*}_{t}$ are the corresponding pseudo process dynamics and pseudo process noise respectively. The term $\mathcal{H} y_t$ is considered as a known input and is approximated with $\mathcal{H} \hat{y}_{t}$. The pseudo gain term $\mathcal{H}$ is a design parameter chosen such that the cross-correlation between the pseudo process noise and the sensor noise is made zero. That is,
\begin{equation*}
    \bbe\left[w^{*}_{t} v^{\top}_{t} \right] = 0 \\
    \implies \bbe \left[(w_{t} - \mathcal{H} v_t) v^{\top}_{t} \right] = M  - \mathcal{H} \Sigma_{v} = 0, \\
    \implies \mathcal{H} = M \Sigma^{-1}_{v}.
\end{equation*}
The mean and the covariance of the pseudo process noise are then given by
\begin{align}
     \bbe \left[w^{*}_{t} \right] 
     &= \bbe \left[w_{t} - \mathcal{H} v_t \right] = \bbe \left[w_{t}\right] - \mathcal{H} \bbe \left[v_t \right]= 0, \\
    \Sigma^{*}_{w} 
    &= \bbe \left[w^{*}_{t} {w^{*}_{t}}^{\top} \right] \nonumber\\
    &= \bbe \left[(w_{t} - \mathcal{H} v_{t}) (w_{t} - \mathcal{H} v_{t})^{\top} \right] \nonumber\\
    &= \bbe \left[ w_{t} w^{\top}_{t} - w_{t} v^{\top}_{t} \mathcal{H}^{\top} - \mathcal{H} v_{t} w^{\top}_{t} + \mathcal{H} v_{t} v^{\top}_{t} \mathcal{H}^{\top} \right] \nonumber\\
    &= \Sigma_{w} - M \mathcal{H}^{\top} - \mathcal{H} M^{\top} + \mathcal{H} \Sigma_{v} \mathcal{H}^{\top}\nonumber\\
    &= \Sigma_{w} - M \Sigma^{-1}_{v} M^{\top} - \mathcal{H} \Sigma_{v} \mathcal{H}^{\top} + \mathcal{H} \Sigma_{v} \mathcal{H}^{\top}\nonumber\\
    &= \Sigma_{w} - M \Sigma^{-1}_{v} M^{\top}.
\end{align}
Then, applying Schur complement on \eqref{eqn_w_v_cross_correl}, it is easy to observe that $\Sigma^{*}_{w}$ is positive definite as well:
\begin{equation*}
    \begin{bmatrix} \Sigma_{w} & M \\ M^{\top} & \Sigma_{v} \end{bmatrix} \succ 0 \iff \Sigma_{v} \succ 0, \, \underbrace{\Sigma_{w} - M \Sigma^{-1}_{v} M^{\top}}_{\Sigma^{*}_{w}} \succ 0.
\end{equation*}
With the new pseudo process equation $f^{*}(\cdot)$ given by \eqref{eqn_pseudo_env_dyn} and the measurement update given by \eqref{eqn_output_sensor_model}, the standard Unscented Kalman Filter can be implemented as the new process noise $w^{*}_{t}$ is no longer cross-correlated with the sensor noise $v_{t}$.

\textbf{The Unscented Kalman Filter.} The UKF is a popular tool for nonlinear state estimation. It uses the so-called Unscented Transformation (UT) in both the propagation and update step, yielding derivative-free Kalman filtering for nonlinear systems. For Gaussian inputs, the moment estimates from UT are accurate up to the third order approximation and for the case of non-Gaussian, the approximations are accurate to at least the second-order as described in \cite{wan2000unscented}. An ensemble of $2 n_z + 1$ samples called the \emph{sigma points} following the Van Der Merwe algorithm is generated deterministically as follows:
\begin{align}
    \chi_{0} &= \hat{Z}_{t-1} \\
    \chi_i &= \hat{Z}_{t-1} + \left[ \sqrt{(n_z + \lambda) \hat{\Sigma}_{\mathcal{Z}_{t-1}}} \right]_{i}, i = [1:n_z] \\
    \chi_{i+n} &= \hat{Z}_{t-1} - \left[ \sqrt{(n_z + \lambda) \hat{\Sigma}_{\mathcal{Z}_{t-1}} } \right]_{i}, i = [1:n_z],
\end{align}
where $\hat{Z}_{t-1}, \hat{\Sigma}_{\mathcal{Z}_{t-1}}$ denote the posterior UKF estimate of mean and covariance of the environmental state at $t-1$ and $\left[ \sqrt{(n_z + \lambda)\hat{\Sigma}_{\mathcal{Z}_{t-1}}} \right]_{i}$ is the $i^{th}$ row or column of the matrix square root $ \sqrt{(n_z + \lambda)\hat{\Sigma}_{\mathcal{Z}_{t-1}}}$. The weights of the sigma points in the calculation of propagated mean and covariance are 
\begin{align}
    W^{(m)}_{0} &= \frac{\lambda}{n_z + \lambda},\ W^{(c)}_{0} = \frac{\lambda}{n_z + \lambda} + 1 - \alpha^2_u + \beta_{u}, \\
    W^{(m)}_{i} &= W^{(c)}_{i} = \frac{\lambda}{2(n_z + \lambda)},\ i = 1, 2, \dots,2n_z.
\end{align}
The scaling parameter $\lambda = \alpha^2_u (n + \kappa) - n_z$, where $\alpha_u, \beta_{u}, \kappa$ are used to tune the unscented transformation. Then, every sigma point is propagated through non-linear state update equation to yield an ensemble of sigma points capturing the \textit{a priori} statistics of $\mathcal{Z}_{t}$ namely $\mathcal{Z}^{-}_{t}, \Sigma^{-}_{\mathcal{Z}_{t}}$. In the update step, we use the transformed set of sigma points described above and propagate them using the given nonlinear measurement model. Finally, the aposteriori state $\hat{Z}_{t}$ and covariance $\hat{\Sigma}_{\mathcal{Z}_{t}}$ are computed using the output residual and the obtained Unscented Kalman filter gain, $\mathcal{L}_{t}$ as follows
\begin{align} \label{eqn_ukf}
    \xi_i &= f^{*}(\chi_i, u_t), \quad i = 0,1,\dots,2n_z, \\
   \mathcal{Z}^{-}_{t} &= \sum^{2n_z}_{i=0}W^{(m)}_{i} \xi_i, \\ 
    \Sigma^{-}_{\mathcal{Z}_{t}} &= \sum^{2n_z}_{i=0}W^{(c)}_{i} (\xi_i - \mathcal{Z}^{-}_{t})(\xi_i - \mathcal{Z}^{-}_{t})^{\top} + \Sigma^{*}_{w} \\
    \Theta_i &= \mathcal{S}(\xi_i), \quad i = 0,1,\dots,2n_z, \\
    \mu_{\Theta} &= \sum^{2n_z}_{i=0}W^{(m)}_{i} \Theta_i, \\
    \Sigma_{\Theta} &= \sum^{2n_z}_{i=0}W^{(c)}_{i} (\Theta_i - \mu_{\Theta})(\Theta_i - \mu_{\Theta})^{\top} + \Sigma_{v} \\
    \mathcal{L}_{t} &= \left[\sum^{2n_z}_{i=0}W^{(c)}_{i} (\xi_i - \mathcal{Z}^{-}_{t})(\Theta_i - \mu_{\Theta})^{\top}\right] \Sigma^{-1}_{\Theta} \\
    \hat{Z}_{t} &= \mathcal{Z}^{-}_{t} + \mathcal{L}_{t} (y_t-\mu_{\Theta}) \\
    \hat{\Sigma}_{\mathcal{Z}_{t}} &= \Sigma^{-}_{\mathcal{Z}_{t}} - \mathcal{L}_{t} \Sigma_{\Theta} \mathcal{L}^{\top}_{t}.
\end{align}

\paragraph*{\bf Remark} While the values of the scaling parameters $\lambda, \alpha_{u}, \beta_{u}, \kappa$ that yield third-order approximation accuracy for Gaussian state distributions are well known, there are still no known guidelines to chose the best values for the above parameters when the state distributions are non-Gaussian. With the environmental state distribution being generally non-Gaussian and along with the moments calculation being done using finite number of deterministic samples result in the obtained moments being only approximations of the true moments of the non-Gaussian state distributions. This motivates our use of distributionally robust risk constraints to explicitly account for this non-Gaussianity.

\subsection{Nonlinear Model Predictive Control (NMPC) Law} \label{sec:NLMPC}
\begin{figure}
    \centering
    \includegraphics[scale=0.5]{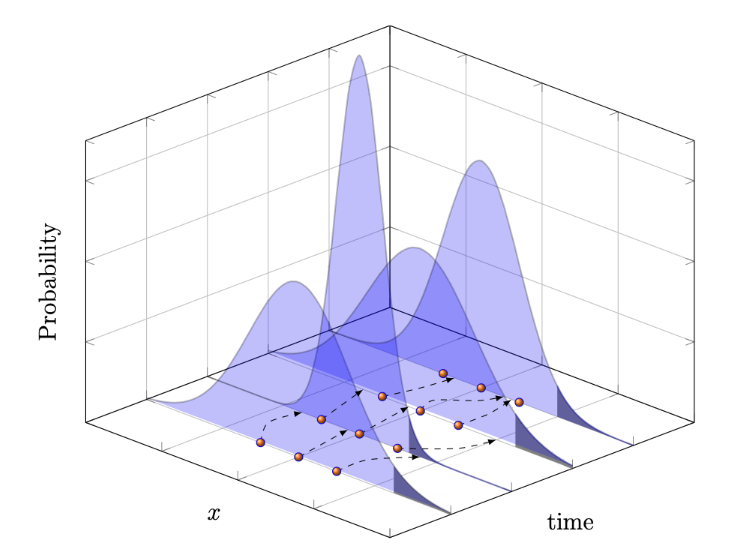}
\caption{An output feedback based controller using NMPC and UKF is illustrated here in $\bbr$. The sigma points and the distribution of the states at different time steps are depicted with orange dots and shaded blue area respectively. The NMPC-steered trajectory from each sigma point is depicted in dotted black lines. At each time step $t$, the distributionally robust budget risk constraint is shown using darkly shaded color whose area is at most $\alpha_{t}$.}
    \label{fig_nmpc_ukf}
\end{figure}

In this subsection, we elaborate about the feedback controller that represents an important module in the autonomy stack shown in Figure \ref{fig_dr_autonomy_stack}. Since the robot dynamics is both nonlinear and stochastic, we will be describing an output feedback based nonlinear model predictive control algorithm to steer the robots in the considered setting. Specifically, the steering law for steering the robots is realized through a multiple shooting-based nonlinear model predictive controller \cite{gu2005}, \cite{Allgower_NMPC} that uses the state estimate obtained through an Unscented Kalman filter at each time step. An illustration of the output feedback is shown in Figure~\ref{fig_nmpc_ukf}.

The error at a time step $t$ is 
\begin{align} \label{eqn_error_t}
    e_{t} = C_{xr}\hat{Z}_{t} - x_{s} = \hat{x}_{t} - x_{s},
\end{align}
where $x_s$ represents a sample in the free space to be steered to. At every time instance $t$, the nonlinear model predictive controller repeatedly solves the following optimization problem in a receding horizon fashion with prediction horizon $N_{t}$ to find a control input sequence as follows
\begin{align} \label{eqn_nmpc_defn}
    u^{\dagger}_{t} = \argmin_{\left\{u_{k}\right\}^{t+N_{t}-1}_{k = t}} \quad & \sum^{t+N_{t}-1}_{k=t} \left( \left \| e_{k} \right \Vert^{2}_{Q} + \left \| u_{k} \right \Vert^{2}_{R} \right) + \left \| e_{k+N_{t}} \right \Vert^{2}_{Q}, \\
    \text{subject to} \quad & \hat{x}_{k+1} = \mathbf{f}(\hat{x}_k, u_k), \nonumber\\
    & \hat{x}_{k} \in \mathcal{X}, \ u_{k} \in \mathcal{U}, \nonumber
\end{align}
where only the first control input in the optimal sequence $u^{\dagger}_{t}$ is applied at time $t$ and the horizon is shifted to $t+1$ for the problem to be solved again. Here, $Q, R$ denote the state penalty and control penalty matrices respectively.

\section{Sampling-Based Motion Planning Algorithm} \label{sec_drrrt_algorithm}
The planner module of an autonomy stack as shown in the Figure \ref{fig_dr_autonomy_stack} is responsible for planning a path from a given source to a desired destination adhering to all motion planning specifications. Typically, the planner module solves a motion planning problem as in \eqref{eqn_dr_motion_planning_new} to design the reference trajectory from source to destination.
However, due to the non-convex collision avoidance constraints and presence of uncertainties, it is difficult to exactly solve the motion planning problem subject to the distributionally robust risk constraints. Hence, we resort to a sampling based motion planner,  which incrementally constructs a motion plan from the source to the destination for a robot by sampling a point in the obstacle free space and connecting the point to the tree of motion plans if the trajectory connecting the points is collision-free. We propose to use a distributionally robust, kinodynamic variant of the $\rrtstar$ motion planning algorithm with dynamic output feedback policies. Our proposed algorithm, called Nonlinear Risk Bounded $\rrtstar (\nrbrrtstar)$, grows trees of state and state estimate \emph{distributions}, rather than merely trees of states, and incorporates distributionally robust risk constraints to build risk-bounded state trajectories and feedback policies. The $\nrbrrtstar$ tree expansion procedure, inspired by the chance constrained $\rrtstar$ algorithm developed in \cite{luders_rrtstar}, is presented in Algorithm \ref{alg_tree_expansion}. The $\nrbrrtstar$ tree is denoted by $\mathcal{T}$, consisting of $|\mathcal{T}|$ nodes. Each node $N$ of the tree $\mathcal{T}$ consists of a sequence of state distributions, characterized by a distribution mean $\hat{x}$ and covariance $D$. A sequence of means and covariance matrices is denoted by $\bar{\sigma}$ and $\bar{\Pi}$, respectively. The final mean and covariance of a node's sequence are denoted by $x[N]$ and $D[N]$, respectively. For the state distribution sequence $(\bar{\sigma},\bar{\Pi})$, the notation $\Delta J(\bar{\sigma},\bar{\Pi})$ denotes the cost of that sequence. If $(\bar{\sigma},\bar{\Pi})$ denotes the trajectory of node $N$ with parent $N_{parent}$, then we denote by $J[N]$, the entire path cost from the starting state to the terminal state of node $N$, constructed recursively as
\begin{align} \label{eqn_node_cost}
    J[N] = J[N_{parent}] + \Delta J(\bar{\sigma},\bar{\Pi}).
\end{align}

\subsection{Approximating Optimal Cost-To-Go}
\begin{figure}[H]
    \centering
    \includegraphics[scale=0.4]{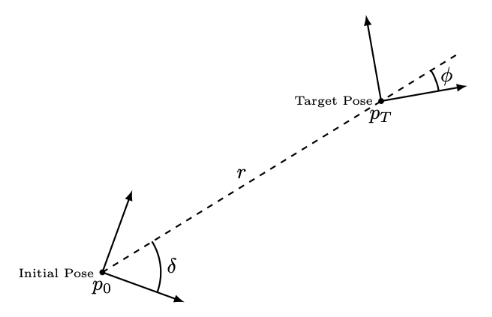}
\caption{The problem of steering a non-holonomic robot from an initial pose $p_{0}$ to a target pose $p_{T}$ with different orientation is shown here. }
    \label{fig_egocentric}
\end{figure}

In order to efficiently explore the reachable set of the dynamics and increase the likelihood of generating collision-free trajectories, authors in \cite{frazzoli_real_mp} described the limitations of the standard Euclidean distance metric and rather advocated the use of an optimal ``cost-to-go'' metric, that takes into account a dynamics and control-related quantities  to steer the robot. When the dynamics are linear and obstacles are ignored, the optimal ``cost-to-go'' between any two nodes in the tree can be computed using dynamic programming. However, with nonlinear dynamics and potentially non-holonomic constraints, cost-to-go can be tailored to the specific robot dynamics. Inspired by \cite{park_nonhol_dist}, we present an approximate cost-to-go for some special robot representations and car-like dynamics. The initial pose of a robot, $p_{0}$ can be defined using a tuple $(x, y, \psi)$, where $(x,y)$ are the positions in global coordinate frame and $\psi$ being its orientation with respect to the positive $x$ axis. We use an egocentric polar coordinate system to describes the relative location of the target pose $p_{T}$ observed from the robot with radial distance $r$ along the line of sight vector from the robot to the target, orientation of the target $\phi$, and orientation of the robot $\delta$, where angles are measured from the line of sight vector as shown in the Figure~\ref{fig_egocentric}.

\begin{definition}
The non-holonomic directed distance from an initial pose $p_{0}$ to a target pose $p_{T}$, with non-holonomic constraints is defined as
\begin{align} \label{eqn_nonhol_dist}
    \mathfrak{D}(p_{0},p_{T}) = \sqrt{r^2 + k^{2}_{\phi} \phi^2} + k_{\delta} | \delta |. 
\end{align}
\end{definition}
Here, $k_{\phi} > 0$ is a constant that represent the weight of $\phi$ with respect to the radial distance $r$ and $k_{\delta} > 0$ is another constant that emphasizes the weight on $|\delta|$ as $\delta$ can take both positive and negative values. The non-holonomic distance in \eqref{eqn_nonhol_dist} is asymmetric meaning that $\mathfrak{D}(p_{0},p_{T})$ is not equal to $\mathfrak{D}(p_{T},p_{0})$.



\subsection{Outline of $\nrbrrtstar$ Algorithm}
In the first step, using the \texttt{Sample$(\cdot)$} function, a random pose $x_{rand}$ is sampled from the free space $\mathcal{X}^{\texttt{free}}_{t}$. Then the tree node, $N_{nearest}$ that is nearest to the sampled pose is selected using the \texttt{NearestNode$(\cdot)$} function (line 3 of Algorithm \ref{alg_tree_expansion}) which uses the ``cost-to-go'' distance metric defined in \eqref{eqn_nonhol_dist}. Attempts are then made to steer the robot from the nearest tree node to the random sample using the \texttt{Steer$(\cdot)$} function that employs the steering law explained in Section \ref{sec_steer} (line 4). The control policy obtained is then used to propagate the state mean and covariance, and the entire trajectory $(\bar{\sigma},\bar{\Pi})$ is returned by the \texttt{Steer$(\cdot)$} function. Using the DR-Feasible$(\cdot)$ function, each state distribution in the trajectory is checked for distributionally robust probabilistic constraint satisfaction discussed in the next section. Further, the line connecting subsequent state distributions in the trajectory are also checked for collision with the obstacle sets $\mathcal{O}_{it}, i \in \mathcal{B}$. An outline of the DR-Feasible subroutine is shown in Algorithm \ref{alg_dr_feasible_module}. If the entire trajectory $(\bar{\sigma},\bar{\Pi})$ is probabilistically feasible, a new node $N_{min}$ with that distribution sequence $(\bar{\sigma},\bar{\Pi})$ is created (line 7) but not yet added to $\mathcal{T}$. Instead, nearby nodes are identified for possible connections via the \texttt{NearNodes$(\cdot)$} function (line 8), which returns a subset of nodes $\mathcal{N}_{near} \subseteq \mathcal{T}$, if they are within a search radius ensuring probabilistic asymptotic optimality guarantees specified in \cite{revisiting_rrtstar}. The search radius is given by
\begin{align}
    r = \min\left\{ \gamma \left( \frac{\log(|\mathcal{N}_{t}|)}{\mathcal{N}_{t}}  \right)^{\frac{1}{d+1}}, \mu_{max} \right\}, 
\end{align}
    where $\mathcal{N}_{t}$ refers to the number of nodes in the tree at time $t$, the positive scalar $\mu_{max}$ is the maximum radius specified by the user, and $\gamma$ refers to the planning constant based on the $d$ dimensional environment and is selected using \cite[Theorem 1]{revisiting_rrtstar}. Then we seek to identify the lowest-cost, probabilistically feasible connection from the $\mathcal{N}_{near}$ nodes to $x_{rand}$ (lines 10-14). For each possible connection, a distribution sequence is simulated via the steering law (line 11). If the resulting sequence is probabilistically feasible, and the cost of that node represented as $c_{rand} = J[N_{near}] + \Delta J(\bar{\sigma},\bar{\Pi}$, is lower than the cost of $N_{min}$ denote by $J[N_{min}]$, then a new node with this sequence replaces $N_{min}$ (line 14). The lowest-cost node is ultimately added to $\mathcal{T}$ (line 15). Finally, edges are rewired based on attempted connections from the new node $N_{min}$ to nearby nodes $\mathcal{N}_{near}$ (lines 17-22), ancestors excluded (line 17) which are found using \texttt{Ancestors$(\cdot)$} function. A distribution sequence is simulated via the steering law from $N_{min}$ to the terminal state of each nearby node $N_{near} \in \mathcal{N}_{near}$ (line 18). If the resulting sequence is probabilistically feasible, and the cost of that node $c_{min}$ is lower than the cost of $N_{near}$ given by $J[N_{near}]$ (line 19), then a new node with this distribution sequence replaces $N_{near}$ (lines 21-22). The tree expansion procedure is then repeated until a node from the goal set is added to the tree. At that point, a distributionally robust feasible trajectory is obtained from the tree root to $\mathcal{X}_{goal}$.

\begin{algorithm}
\caption{\texttt{$\nrbrrtstar$}- Tree Expansion Procedure} \label{alg_tree_expansion}
\begin{algorithmic}[1]
\State $\text{Inputs: Current Tree } \mathcal{T}, \text{ time step } t$
\State $x_{rand} \gets$ Sample$(\mathcal{X}^{\texttt{free}}_t)$
\State $N_{nearest} \gets$ NearestNode$(x_{rand}, \mathcal{T})$
\State $(\bar{\sigma}, \bar{\Pi}) \gets$ \texttt{Steer}$(\hat{x}[N_{nearest}], D[N_{nearest}], x_{rand})$
\State \text{\color{gray} // \texttt{Check if sequence} $(\bar{\sigma}, \bar{\Pi})$ \texttt{is DR-Feasible}}
\If{DR-Feasible$(\bar{\sigma}, \bar{\Pi})$} 
\State Create node $N_{min}\{\bar{\sigma}, \bar{\Pi}\}$
\State $\mathcal{N}_{near} \gets$ NearNodes$(\mathcal{T}, x_{rand}, \left | \mathcal{T} \right \vert)$
\State \text{\color{gray} // \texttt{Connect via a minimum-cost path}}
\ForEach {$N_{near} \in \mathcal{N}_{near} \backslash N_{nearest}$}
\State $(\bar{\sigma}, \bar{\Pi}) \gets$ \texttt{Steer}$(\hat{x}[N_{near}], D[N_{near}], x_{rand})$
\State $c_{rand} \gets$ J$[x_{near}] + \Delta J(\bar{\sigma}, \bar{\Pi})$
\If{DR-Feasible$(\bar{\sigma}, \bar{\Pi})$ \& $c_{rand} < J[N_{min}]$}
\State Replace $N_{min}$ with $N_{min}\{\bar{\sigma}, \bar{\Pi}\}$
\EndIf
\EndFor
\State Add $N_{min}$ to $\mathcal{T}$
\State \text{\color{gray} // \texttt{Re-Wire the Tree}}
\ForEach {$N_{near} \in \mathcal{N}_{near} \backslash$ Ancestors$(N_{min})$}
\State $(\bar{\sigma}, \bar{\Pi}) \gets$ \texttt{Steer}$(\hat{x}[N_{min}], D[N_{min}], \hat{x}[N_{near}])$
\State $c_{min} \gets$ J$[N_{min}] + \Delta J(\bar{\sigma}, \bar{\Pi})$
\If{DR-Feasible$(\bar{\sigma}, \bar{\Pi})$ \& $c_{min} <  J[N_{near}]$}
\State Delete $N_{near}$ from $\mathcal{T}$
\State Add new node $N_{new}\{\bar{\sigma}, \bar{\Pi}\}$ to $\mathcal{T}$
\EndIf
\EndFor
\EndIf
\end{algorithmic}
\end{algorithm}

\begin{algorithm}
\caption{DR-Feasible Subroutine}\label{alg_dr_feasible_module}
\begin{algorithmic}[2]
\State $\text{Input: } \mathbf{T-} \text{time step distribution sequence} (\bar{\sigma}, \bar{\Pi})$
\For{$t = 1$ to $\mathbf{T}$}
\State $(\hat{x}_t, D_{\hat{x}_{t}}) \gets$ $t^{th}$ element in $(\bar{\sigma}, \bar{\Pi})$ sequence
\State $\mathbb{L} \gets$ Line connecting position block of $\hat{x}_{t-1}$ to
$\hat{x}_{t}$
\ForEach{$i \in \mathcal{B}$}
\If{$(\hat{x}_t, D_{\hat{x}_{t}})$ dissatisfies \eqref{eqn_dr_constraint_tightening} or $\mathbb{L} \cap \mathcal{O}_{it} \neq \varnothing$}
\State Return \texttt{false} 
\EndIf
\EndFor
\EndFor
\State Return \texttt{true}
\end{algorithmic}
\end{algorithm}

\section{Distributionally Robust Collision Check}
\label{sec_dr_collision_check}
It is necessary to ensure that the total risk of the trajectory returned by the planner module of the autonomy stack in Figure \ref{fig_dr_autonomy_stack} does not exceed the given total risk budget and thereby agree with the motion planning specifications. In this section, we evaluate the safety of a trajectory returned by the planner module using distributionally robust risk constraints and formally present a risk treatment to ensure the safety of the whole mission plan. 

\subsection{Moment-Based Ambiguity Set To Model Uncertainty}
Unlike most stochastic motion planning algorithms that often assume a functional form (often Gaussian) for probability distributions to model uncertainties, we will focus here on uncertainty modeling using moment-based ambiguity sets. Since the initial environmental state $\mathcal{Z}_0$ is not assumed to be Gaussian, then neither are the environmental state distributions $\bbp_{\mathcal{Z}_{t}}$ at any point of future time $t$. The ambiguity set defining the true environmental state is
\begin{equation} \label{eqn_ambiguity_set_env_state}
     \mathcal{P}^{\mathcal{Z}_{t}} = \left\{ \bbp_{\mathcal{Z}_t} \mid \bbe[\mathcal{Z}_t] = \hat{\mathcal{Z}}_{t}, \bbe [(\mathcal{Z}_t - \hat{\mathcal{Z}}_t)(\mathcal{Z}_t - \hat{\mathcal{Z}}_t)^{\top}] = \Sigma_{\mathcal{Z}_t} \right\}.
\end{equation}

\subsection{Risk Treatment for Trajectory Safety} \label{subsec_risk_treatment}
To study the stage risk constraints \eqref{eqn_dr_motion_planning_new} for the safety of the planned reference trajectory, we follow the steps similar to the one described in \cite{safaoui2021risk}. Let $\mathfrak{P}_{Safe}$ denote the event that plan $\mathfrak{P}$ succeeds and $\mathfrak{P}_{Fail}$ as the complementary event (i.e. failure). Given a confidence $\beta \in (0,0.5]$, we consider the specification that a plan succeeds with high probability or equivalently that it fails with low probability. That is, 
\begin{align}
 \bbp(\mathfrak{P}_{Safe}) \geq 1 - \beta \iff \bbp(\mathfrak{P}_{Fail}) \leq \beta.  
\end{align}
Failure of the total plan requires at least one stage risk constraints to be violated. Then, $\forall t \in [0:T]$,
\begin{align*}
    & \underset{\bbp_{\mathcal{Z}_t}\in \mathcal{P}^{\mathcal{Z}}}{\inf} \bbp_{\mathcal{Z}_t} (C_{xr} \mathcal{Z}_{t} \in \mathcal{X}^{\texttt{free}}_{t}) \geq 1 - \alpha_{t} \\
    \Leftrightarrow & \underset{\bbp_{\mathcal{Z}_t}\in \mathcal{P}^{\mathcal{Z}}}{\sup} \bbp_{\mathcal{Z}_t} (C_{xr} \mathcal{Z}_{t} \notin \mathcal{X}^{\texttt{free}}_{t}) \leq \alpha_{t}.
\end{align*}
Applying Boole's law, probability of the success event can be bounded as 
\begin{align*}
    \bbp(\mathfrak{P}_{Safe}) = 1 - \bbp(\mathfrak{P}_{Fail}) &= 1 - \bbp_{\mathcal{Z}_t} \rlpar{ \bigcup_{t=0}^T C_{xr} \mathcal{Z}_{t} \not\in \mathcal{X}^{\texttt{free}}_{t}} \\
    &\geq 1 - \underset{\bbp_{\mathcal{Z}_t}\in \mathcal{P}^{\mathcal{Z}}}{\sup} \bbp_{\mathcal{Z}_t} \rlpar{ \bigcup_{t=0}^T C_{xr} \mathcal{Z}_{t} \not\in \mathcal{X}^{\texttt{free}}_{t}} \\
    &\geq 1 - \sum_{t=0}^T \underset{\bbp_{\mathcal{Z}_t}\in \mathcal{P}^{\mathcal{Z}}}{\sup} \bbp_{\mathcal{Z}_t} \rlpar{ C_{xr} \mathcal{Z}_{t} \not\in \mathcal{X}^{\texttt{free}}_{t}} \\
    &\geq 1 - \underbrace{\sum_{t=0}^T \alpha_t}_{:=\beta}
\end{align*}
If the stage risks $\forall t \in [0:T]$ are equal, meaning $\alpha_t = \alpha$, then $\beta = (T+1) \alpha$. Furthermore, if the stage risk $\alpha_t = \alpha$ is equally distributed over all $n_{total}$ constraints, then, a risk bound for a single constraint is $\frac{\alpha}{n_{total}}$. Subsequently, the corresponding risk bound for an obstacle $i \in \mathcal{B}$ and the environment $\calx$ would be $\frac{\alpha n_{ob_i}}{n_{total}}$ and $\frac{\alpha n_{env}}{n_{total}}$ respectively. It is possible to have a spatio-temporal based risk allocation and is currently being pursued as a future work.

\subsection{Distributionally Robust Collision Check}
The non-convex obstacle avoidance constraints for obstacle $i \in \mathcal{B}$ can be expressed as the disjunction
\begin{align}
    \neg (A_{i} C_{xr} \mathcal{Z}_{t} \leq b_{it}) \Leftrightarrow \bigvee^{n_i}_{j=1} (a^{\top}_{ij} C_{xr} \mathcal{Z}_{t} \geq a^{\top}_{ij} c_{ijt}),
\end{align}
where $\lor$ denotes disjunction (logical OR operator) and $\neg$ denotes negation (logical NOT operator) and $c_{ijt} = \hat{c}_{ijt} + r_{i,t}$ is a point
nominally on the $j$th constraint of the $i$th obstacle whose covariance is the same as that of $r_{i,t}$. The control law returned by the steering function should also satisfy the state constraints which are expressed as distributionally robust chance constraints. The nominal state constraints, $C_{xr} \mathcal{Z}_t \in \mathcal{X}^{\texttt{free}}_{t}$, are required to be satisfied with probability $1 - \alpha$, under the worst case probability distribution in the ambiguity set. Let the mean and covariance of the uncertain obstacle motion be defined using $\mathbf{E}[c_{ijt}] = \hat{c}_{ijt}$ and $\mathbf{E}[(c_{ijt} - \hat{c}_{ijt})(c_{ijt} - \hat{c}_{ijt})^{\top}] = \Sigma^{c}_{jt}$. Following \cite{drcclp}, under the moment-based ambiguity set defined by \eqref{eqn_ambiguity_set_env_state}, a constraint on the worst-case probability of violating the $j^{th}$ constraint of obstacle $i \in \mathcal{B}$
\begin{equation} \label{eqn_dr_constraint}
    \underset{\bbp_{\mathcal{Z}_t} \in \mathcal{P}^{\mathcal{Z}}}{\sup} \bbp_{\mathcal{Z}_t}(a^{\top}_{ij} C_{xr} \mathcal{Z}_t \geq a^{\top}_{ij} c_{ijt}) \leq \alpha_{i}
\end{equation}
is equivalent to the linear constraint on the state mean $\hat{\mathcal{Z}}_t$
\begin{equation} \label{eqn_dr_constraint_tightening}
    a^{\top}_{ij} C_{xr} \hat{\mathcal{Z}}_t \geq a^{\top}_{ij} \hat{c}_{ijt} + \sqrt{\frac{1 - \alpha_i}{\alpha_i}} {\left \| (D_{\hat{x}_{t}} + \Sigma^{c}_{jt})^{\frac{1}{2}} a_{ij} \right \Vert}_{2},
\end{equation}
where $D_{\hat{x}_{t}} = C_{xr} \Sigma_{\mathcal{Z}_{t}} C^{\top}_{xr}$ and $\alpha_{i}$ is the user prescribed risk parameter for obstacle $i \in \mathcal{B}$. Obstacle risks are allocated such that their sum does not exceed the total constraint risk $\alpha$ (as described in subsection \ref{subsec_risk_treatment}). 
The scaling constant $\sqrt{\frac{1 - \alpha_i}{\alpha_i}}$ in the deterministic tightening of the nominal constraint in \eqref{eqn_dr_constraint_tightening} is larger than the one obtained with a Gaussian assumption, leading to a stronger tightening that reflects the weaker assumptions about the uncertainty distributions. 

\section{Simulation Results} \label{sec_sim_results}
In this section, we demonstrate our proposed framework using simulations with both a unicycle and a bicycle robot dynamics. Specifically, we demonstrate the bicycle dynamics in an urban driving simulator \cite{carla_sim}. The robot is assumed to be moving in a bounded and cluttered environment. While the proposed framework can handle both dynamic and uncertain obstacles, we assume the obstacles to be static and deterministic $(w_{\mathcal{O}_{t}}=0, \forall t)$ for simplicity, so that all uncertainty comes from the unknown initial state, robot process disturbance, and measurement noise. That is, we assume that $g_i(x) = 0, \forall i \in \mathcal{B}$.
\subsection{Unicycle Model Based Simulation}
\subsubsection{Robot Motion Model:} 

\begin{figure}
    \centering
    \includegraphics[scale=0.5]{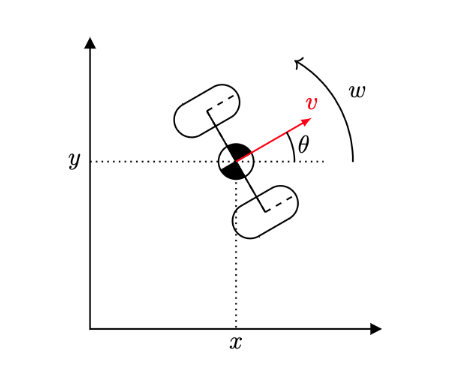}
\caption{A unicycle robot operating in $\bbr^{2}$ is shown here.}
    \label{fig_unicycle}
\end{figure}

We consider the problem of navigating a robot with unicycle dynamics from an initial state to a final set of states. The nonlinear robot and the static obstacle dynamics as shown in Figure \ref{fig_unicycle} are given by
\begin{align} \label{eqn_st_unicycle}
\begin{bmatrix}x_{t+1} \\
y_{t+1} \\ \theta_{t+1} \\ x_{obs, t+1} \\ y_{obs, t+1} \end{bmatrix} &= \underbrace{\begin{bmatrix}x_{t} \\
y_{t} \\ \theta_{t} \\ x_{obs, t} \\ y_{obs, t} \end{bmatrix}}_{\mathcal{Z}_{t}} + \Delta t \begin{bmatrix} \nu_{t}\cos(\theta_{t}) \\ \nu_{t}\sin(\theta_{t}) \\ \omega_{t} \\ 0 \\ 0 \end{bmatrix} + \Delta t \, w_{t} 
\end{align}
where $x_{t}, y_{t} \in \bbr$ are the horizontal and vertical positions of the robot, $\theta_{t} \in \bbr$ is the heading of the robot relative to the $x$-axis with $v_{t}, \omega_{t} \in \bbr$ being the linear and angular velocity control inputs, and $w_{t}$ denoting the disturbances at time $t$. Further, $x_{obs, t}, y_{obs, t} \in \bbr$ denote the the horizontal and vertical positions of the obstacles at time $t$. The discretization time step was selected to be $\Delta t = 0.2$sec.

\subsubsection{Measurement Model:} 
We assume that the unicycle robot is equipped with a sensor with measurement noise. Further we consider independent sensor and process noises for simplicity. The measurement model is given by
\begin{equation} \label{eqn_sensor_model_unicycle}
z_{t} = \mathcal{S}(\mathcal{Z}_{t}) + v_{t} = \mathcal{Z}_{t} + v_{t}.
\end{equation}

\subsubsection{Discussion of Results:} 
\begin{figure}[H]
\begin{center}
    \includegraphics[scale=0.80]{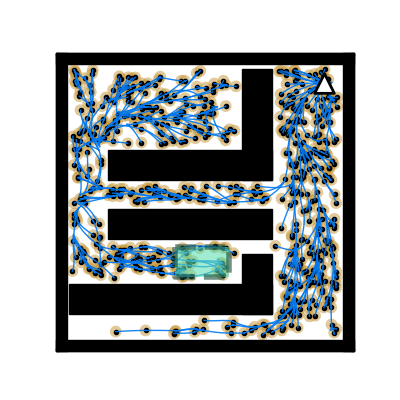}
    \caption{A risk-bounded motion plan of a unicycle robot operating in $\bbr^{2}$ computed using the $\nrbrrtstar$ algorithm with stage-risk budget $\alpha_{t} = 10^{-4}$ and the corresponding $1\sigma$ uncertainty covariance ellipses for its positions are shown here in yellow. The white triangle and the green rectangle respectively denote the start position and the goal region.}
    \label{fig_unicycle_motion_plan}
\end{center}
\end{figure}
\begin{figure*}[!h]
\minipage{0.30\textwidth}
  \includegraphics[scale=0.6]{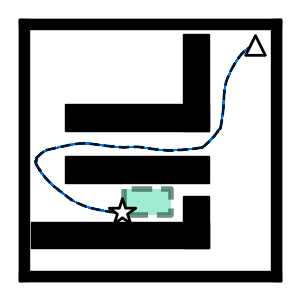}
  \caption{Results of 1000 independent Monte-Carlo trials using zero-mean noises $w_{t}, v_{t}$ sampled from multivariate Laplacian distribution with covariance matrices $\Sigma_{w} = \Sigma_{v} = 10^{-7} I_{n}$ are shown here. With noises being minimal, success rate is $100\%$. The white triangle, star and green rectangle denote the starting position, goal position and the goal region respectively.}
  \label{fig_path_plot_nmpc_lap_1}
\endminipage\hfill
\minipage{0.30\textwidth}
  \includegraphics[scale=0.6]{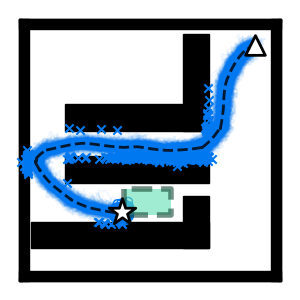}
  \caption{Results of 1000 independent Monte-Carlo trials using zero-mean noises $w_{t}, v_{t}$ sampled from multivariate Laplacian distribution with covariance matrices $\Sigma_{w} = \Sigma_{v} = 10^{-3} I_{n}$ are shown here. With noises being stronger, success rate falls to $44.9\%$. The white triangle, star and green rectangle denote the starting position, goal position and the goal region respectively.}
  \label{fig_path_plot_nmpc_lap_2}
\endminipage\hfill
\minipage{0.30\textwidth}%
  \includegraphics[scale=0.6]{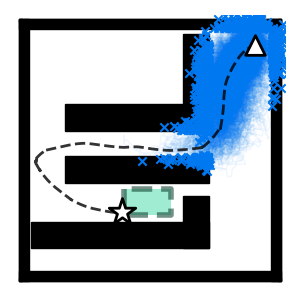}
  \caption{Results of 1000 independent Monte-Carlo trials using zero-mean noises $w_{t}, v_{t}$ sampled from multivariate Laplacian distribution with covariance matrices $\Sigma_{w} = \Sigma_{v} = 10^{-2} I_{n}$ are shown here. With noises being strong, success rate falls to $0\%$. The white triangle, star and green rectangle denote the starting position, goal position and the goal region respectively.}
  \label{fig_path_plot_nmpc_lap_3}
\endminipage
\end{figure*}

The motion plan using $\nrbrrtstar$ algorithm was generated on a machine with an Intel Core i7 CPU and 8GB of RAM. The trajectory tracking Monte Carlo simulations were performed on a machine with a Ryzen 7 2700X and 64GB of RAM. The nonlinear MPC problem in \eqref{eqn_nmpc_defn} is modeled with CasADi Opti and solved with IPOPT solver \cite{casadi}. We demonstrate $\nrbrrtstar$ in a obstacle cluttered environment as shown in Figure \ref{fig_unicycle_motion_plan}. The environment consisted of a root node (white triangle), a goal area (dashed green rectangle), and a $10 \times 10$ environment with rectangular obstacles for its sides (black boundary). The robot is assumed to occupy a single point. The input bounds are $\pm 0.5$ units/sec for linear velocity and $\pm \pi$rad/sec for angular velocity. The $\nrbrrtstar$ steering horizon is $N = 30$ and the NMPC planning horizon is $N_{ll} = 10$. The planner control cost matrix was $R = \text{diag}([1,1])$. The tracking cost matrices were $Q = \text{diag}([100, 100, 10])$ and $R = \text{diag}([1, 1])$ for all $t = [0 : T-1]$, and $Q_{T} = 10 Q$. We used a high-level plan risk bound of $\beta = 0.1$. It is divided equally across the time steps $T_{max} = 1000$ and among the obstacle constraints. The process noise distribution $\bbp_{w}$ and the sensor noise distribution $\bbp_{v}$ were taken as a multivariate Laplace distribution with zero mean and covariance $\Sigma_w = \Sigma_v = 10^{-7} I_{n}$. We employed $\nrbrrtstar$ algorithm with just the euclidean distance metric for generating the motion plan. Clearly, the $\nrbrrtstar$ tree as shown in Figure \ref{fig_unicycle_motion_plan} avoided the unsafe gap to the top-right of the goal region in process of reaching the goal as it was deemed to be too risky. To evaluate the effectiveness of the $\nrbrrtstar$ algorithm, we performed Monte-Carlo simulation using the reference trajectory obtained after 1000 iterations of $\nrbrrtstar$ algorithm. Particularly, at each independent trials out of the total 1000 trials, we realized both the process noise $w_{t}$ and the sensor noise $v_{t}$ from either a multivariate Laplacian distribution or a multivariate Gaussian distribution with the same mean and covariance. The results of the Monte-Carlo trials using multivariate Laplacian noises with covariance matrices $(\Sigma_{w} = \Sigma_{v})$ for different noise levels $(10^{-7} I_{n}, 10^{-3} I_{n}, 10^{-2} I_{n})$ are shown in Figures \ref{fig_path_plot_nmpc_lap_1}, \ref{fig_path_plot_nmpc_lap_2}, \ref{fig_path_plot_nmpc_lap_3} respectively. An analogous results using multivariate Gaussian distribution based noises for same noise levels were also obtained. The Monte-carlo trials subjected to noises from both distributions with different noise levels reveal that as the noise levels were smaller $(\Sigma_{w} = \Sigma_{v} = 10^{-7} I_{n})$, the feedback control layer had good tracking of the reference trajectory provided by the planning layer. However, when the noises got stronger $(\Sigma_{w} = \Sigma_{v} = 10^{-3} I_{n})$, more collisions/failure occurred and finally with even stronger noises, $(\Sigma_{w} = \Sigma_{v} = 10^{-2} I_{n})$, no trial was capable to track the whole trajectory without collision. This demonstrates that as the planner in the autonomy stack plans a reference trajectory with a conservative noise setting, the controller can tolerate noises up to a certain level after which it cannot reach the goal set without collisions. A similar analysis was done in \cite{safaoui2021risk} but without a measurement model with sensor noise. In the next subsection, we simulate a car-like robot having bicycle dynamics with a nonlinear measurement model and sensor noise and demonstrate our approach using an urban driving simulator.

\begin{table}[h]
    \centering
\begin{tabular}{|c|c|c|c|c|c|}
\hline
\multirow{2}{*}{Noise} & \multicolumn{2}{c|}{Multivariate Laplacian} & \multicolumn{2}{c|}{Multivariate Gaussian}\\
\cline{2-5}
 & \# of & Avg. Run & \# of & Avg. Run \\
 $(\times I_{n})$ & collisions & Time (s) & collisions & Time (s)\\
\hline \hline
$10^{-7}$ & 0 & 4.2117 & 0 & 4.2043 \\
\hline
$10^{-3}$ & 551 & 3.5119 & 495 & 3.7024 \\
\hline
$10^{-2}$ & 1000 & 0.8964 & 1000 & 0.9310 \\
\hline
\end{tabular}
\caption{Performance metrics from Monte Carlo trials with different covariance matrices $\Sigma_{w} = \Sigma_{v}$ for noises corresponding to two different distributions are tabulated here. It is evident the multivariate Laplacian being an heavy-tailed distribution results in more collisions than its counterpart multivariate Gaussian.}
    \label{tab:path_table_0p0000006}
\end{table}

\subsection{Bicycle Model Based Simulation}
\subsubsection{Robot Motion Model:} 

\begin{figure}
    \centering
    \includegraphics[scale=0.5]{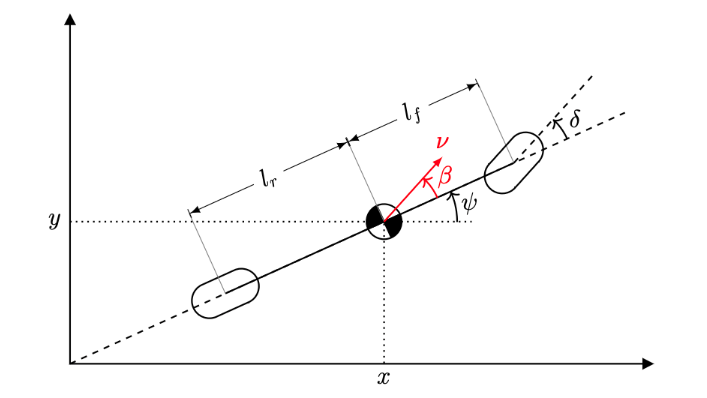}
    \caption{A bicycle model \cite{kongbicycle} depicting the motion of a car-like vehicle is shown here.}
    \label{car_bicycle}
\end{figure}

The kinematics of the robot modeled using bicycle dynamics as shown in Figure~\ref{car_bicycle} and the static obstacle dynamics are given by
\begin{align} \label{eqn_st_bicycle}
\begin{bmatrix} x_{t+1} \\
y_{t+1} \\ \psi_{t+1} \\ \nu_{t+1} \\ x_{obs, t+1} \\ y_{obs, t+1} \end{bmatrix} &= \underbrace{\begin{bmatrix} x_{t} \\
y_{t} \\ \psi_{t} \\ \nu_{t} \\ x_{obs, t} \\ y_{obs, t} \end{bmatrix}}_{\mathcal{Z}_{t}} + \Delta t \begin{bmatrix} \nu_{t} \cos(\psi_{t}) \\ \nu_{t} \sin(\psi_{t}) \\ \frac{\nu_{t}}{L} \tan(\delta_{t}) \\ a_{t} \\ 0 \\ 0 \end{bmatrix} + \Delta t \, w_{t},
\end{align}
where $(x_{t}, y_{t})$ denote the position of center of mass of the robot in the simulation environment, and $\psi_{t}$ its orientation at time $t$. The pair $(x_{obs, t}, y_{obs, t})$ denote the position of center of mass of the obstacle at time $t$. The discretization time step was selected to be $\Delta t = 0.2$. The length of the vehicle wheelbase is denoted by $L$ and $L = l_f + l_r$, where $l_f, l_r$ denote the distance from the center of mass to the front and rear wheels respectively. The angle of the current velocity $\nu_{t}$ of the center of mass with respect to the longitudinal axis of the car is denoted by $\beta_{t}$. The control inputs $(a_{t}, \delta_{t})$ respectively denote the linear acceleration and the steering angle of the robot and acceleration of the center of mass is assumed to be in the same direction as the velocity $\nu_{t}$. We assume the robot to start from the origin with an orientation $(\psi_0 = 270^{\circ}$ in Carla$)$. The pose of the robot is sampled uniformly within the bounds of the feasible environment in $\bbr^{2}$ whose boundaries are not treated probabilistically. We chose to simulate our model using the Audi E-Tron vehicle in the simulator whose parameter values are $L = 2.9m$. 
\subsubsection{Measurement Model:}
We assume that the robot is equipped with a radar sensor which provides a noisy bearing and range to multiple known obstacle locations in the landscape. Thus the measurement model is given by
\begin{equation} \label{eqn_sensor_model_bicycle}
    \footnotesize z_{t} = \mathcal{S}(x_{t}, y_{t}, \theta^{\star}, x_{obs, t}, y_{obs, t}) = \begin{bmatrix} \sqrt{(x_{t} - x_{goal})^2 + (y_{t} - y_{goal})^2} \\ \tan^{-1}\left( \frac{y_{t} - y_{goal}}{x_{t} - x_{goal}} \right) - \psi_{t} \\
    \cos(\theta^{\star}) x_{obs, t} \\
    \sin(\theta^{\star}) y_{obs, t}
    \end{bmatrix} + v_{t},
\end{equation}
where the point $(x_{goal}, y_{goal})$ denotes the location of a landmark inside  the goal region $\mathcal{X}_{goal}$ and $\theta^{\star} = 0.01$ radian denotes a known distortion of a sensor that estimates the position of obstacles. The initial state is assumed to be of zero mean with covariance matrix $\hat{\Sigma}_{\mathcal{Z}_{0}} = 0.001 I_{n_{z}}$. The sensor noise and process noise covariance matrices were $\Sigma_{v} = 0.001I_{p}$ and $\Sigma_{w} = \text{diag}(0.001, 0.001, 0.001, 0.001, 0, 0)$. For simplicity,  we assumed the cross-correlation matrix to be zero $(M = 0)$. While executing the algorithm \ref{alg_tree_expansion}, the equation \eqref{eqn_nonhol_dist} is used to approximate the cost-to-go quantity with $k_{\phi} = 1.2$ and $k_{\delta} = 3$ respectively in the NearestNode$(\cdot)$ and NearNodes$(\cdot)$ modules and for the rest, the cost-to-go is approximated using the steering law that employs nonlinear model predictive control. The search radius used in the Algorithm \ref{alg_tree_expansion} is obtained with $\gamma = 30$. For simplicity, the environmental boundaries are not treated probabilistically. Since the motion model given by \eqref{eqn_st_bicycle} and the measurement model given by \eqref{eqn_sensor_model_bicycle} are nonlinear, this particular set up demands using Unscented Kalman Filter for state estimation and a nonlinear model predictive control for steering. To generate the sigma points and weigh them accordingly, we use $\alpha_u = 1, \beta = 2, \kappa = 3 - n$. Further, we define the error $e_{t} = C_{xr}\hat{\mathcal{Z}}_{t} - x_{s}$ for $i = 0,\dots,T$, with $T=1000$. A dynamic output feedback policy $u_t$ that minimizes the cost function 
\begin{align} \label{eqn_sim_cost_fn}
    J &= \sum^{t+N-1}_{k=t} \left(e^{\top}_{k} Q e_{k} + u^{\top}_{k} R u_{k} \right) + e^{\top}_{k+N} Q e_{k+N} , 
\end{align}
is computed using nonlinear model predictive control to steer the robot from a tree node state $x_t$ to a random feasible sample pose $x_s$ with prediction horizon $N = 50$. The control inputs are constrained as $|a| \leq 3ms^{-2}$ and $|\delta| \leq 70^{\circ}$. The nonlinear model predictive control law is obtained through the CasADi toolbox \cite{casadi} that employs the IPOPT large-scale nonlinear optimization solver. The state and control penalty matrices namely $Q = 100\texttt{diag}([2,2,1,5,1,1]), R = 0.01I_{m}$ are used to penalize the state and control deviations respectively. The distributionally robust state constraints are enforced over a mission horizon length $T=50$ and with probabilistic satisfaction parameter $\alpha = 0.05$. For all the simulations, the $\nrbrrtstar$ algorithm with the distributionally robust collision check or Gaussian collision check procedure is run for 200 iterations, with 2-$\sigma$ position uncertainty ellipses from the covariance matrix being drawn at the end of each trajectory. 
\subsubsection{Results and Discussion:}
\begin{figure}
    \centering
    \includegraphics[scale=0.25]{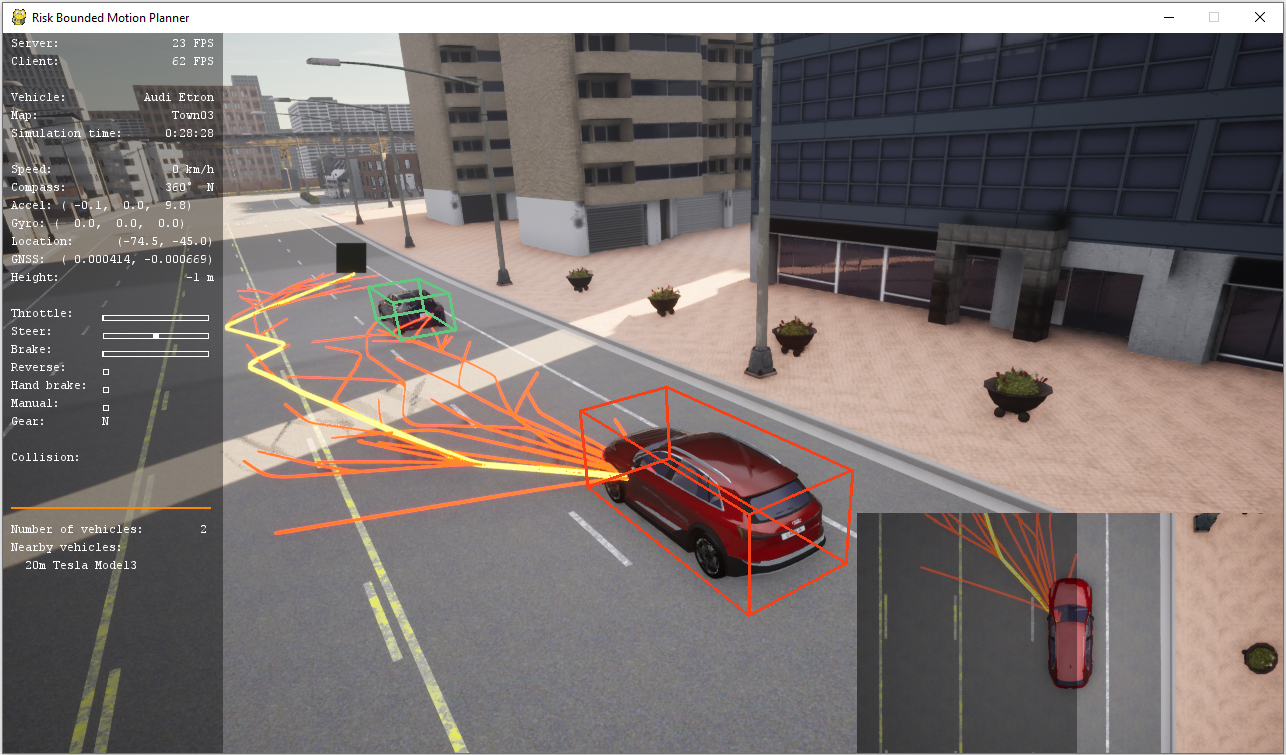}
    \caption{Our proposed approach is demonstrated using an urban driving simulator \cite{carla_sim}. An $\nrbrrtstar$ tree (in orange color) as a result of 200 iterations and the resulting reference path (in yellow color) from the start to the goal region are shown here. The $\nrbrrtstar$ algorithm employed distributionally robust chance constraints with $\alpha = 0.05$.}
    \label{fig_carla_start}
\end{figure}
\begin{figure}[H]
    \centering
    \includegraphics[scale=0.40]{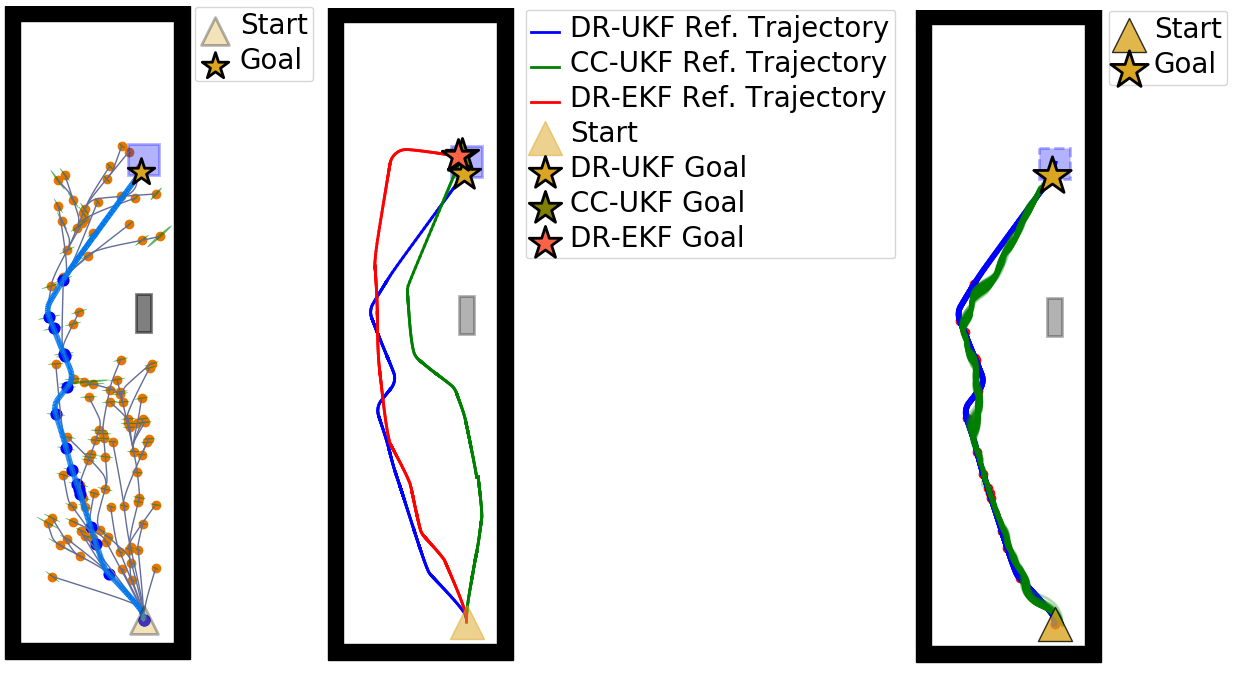}
    \caption{A 2-D projection of the urban driving environment containing the robot starting pose, goal region (blue shaded rectangle), the obstacle (black shaded rectangle), constructed motion plan with distributionally robust chance constraints (in blue color) after executing 200 iterations of $\nrbrrtstar$ algorithm with the two-standard deviation covariance ellipses and the generated reference path from the source to the goal are shown on the plot to the left. A comparison of $\nrbrrtstar$ algorithm with distributionally robust (in blue color(UKF) and red color(EKF)) \& Gaussian chance constraints (in green color(UKF)) are shown on the middle. Clearly, the added conservatism greater than that of the chance constrained counterpart that comes due to the distributional robustness would account for any arbitrary unknown state distribution satisfying the given moments. The results of 100 independent Monte Carlo trials of trajectory tracking with noises drawn from multivariate Gaussian distributions are shown on the right.}
    \label{fig_tree_path}
\end{figure}

The $\nrbrrtstar$ tree expansion procedure was run until at most 2 nodes from the goal region were added to the tree successfully and the resulting reference trajectory from the starting pose to the goal is shown in Figure~\ref{fig_carla_start}. The corresponding 2D projection is shown in the left side of Figure~\ref{fig_tree_path} and distributionally robust risk constraints generated more conservative trajectories around the obstacles, by explicitly incorporating the uncertainty in the state due to the initial localization, system dynamics, and measurement uncertainties in the form of ambiguity sets. It produces trajectories that satisfy the chance constraints under the worst-case distribution in the ambiguity sets. Clearly, the feasible set is smaller with the distributionally robust constraints. On the other hand, a similar reference trajectory was generated using Gaussian chance constraints assuming system uncertainties are Gaussian. The reference trajectory generated using distributionally robust risk constraints (with UKF and EKF), and Gaussian chance constraints are shown in the center of Figure~\ref{fig_tree_path}. The distributionally robust trajectories with a more sophisticated and coherent quantification of risk, are generated with the same computational complexity as with Gaussian chance constraints and exhibit conservatism to account for state distribution ambiguity. Though the result with Extended Kalman Filter based state estimation also produces collision free trajectories, the result inherently suffers from the (first order) approximations due to linearization. To corroborate the results generated by the high-level motion planner, a Monte-Carlo simulation of 100 independent trials involving noises sampled from multivariate Gaussian distribution was conducted and the results are shown in the right side of Figure~\ref{fig_tree_path}. It is evident that given the conservative noise assumption and risk quantification performed using the distributionally robust approach, the realized sampled paths from the source to destination had little deviations around the reference trajectory. This further demonstrates our proposed approach that with a  high-level motion plan that accounts for uncertainty, the low level online tracking controller has more room to maneuver and respond to noise realizations. The video of the simulation is available at \url{https://youtu.be/KpyWXRZ-wSI} and the simulation code is available at \url{https://github.com/TSummersLab/Risk_Bounded_Nonlinear_Robot_Motion_Planning}.


\section{Conclusion} \label{sec_conclusions}
In this paper, we presented a methodological framework aimed towards tighter integration of perception and planning in nonlinear robotic systems. Uncertainties in perception and motion prediction are explicitly accounted for through distributionally robust risk constraints. Using a dynamic output feedback based control policy realized using a nonlinear MPC and an Unscented Kalman filter, a new algorithm called $\nrbrrtstar$ is shown to produce risk bounded trajectories with systematic risk assessment. Potential future research involves using deep learning based perception system with semantic SLAM combined with our approach. We will also explore optimal spatio-temporal risk allocation for obstacles and environmental constraints which can result less conservative motion plans for robots operating in cluttered environments. Another promising future research direction involves studying propagation of state distributions of nonlinear systems with higher order moments and using measure concentration techniques to estimate the risk.

\bibliographystyle{elsarticle-num} 
\bibliography{bibliograph}

\begin{thebibliography}{10}
\expandafter\ifx\csname url\endcsname\relax
  \def\url#1{\texttt{#1}}\fi
\expandafter\ifx\csname urlprefix\endcsname\relax\def\urlprefix{URL }\fi
\expandafter\ifx\csname href\endcsname\relax
  \def\href#1#2{#2} \def\path#1{#1}\fi

\bibitem{blackmore_pioneer}
L.~Blackmore, H.~Li, B.~Williams, A probabilistic approach to optimal robust
  path planning with obstacles, in: 2006 American Control Conference, IEEE,
  2006, pp. 7--pp.

\bibitem{agha_firm}
A.-A. Agha-Mohammadi, S.~Chakravorty, N.~M. Amato, Firm: Sampling-based
  feedback motion-planning under motion uncertainty and imperfect measurements,
  The International Journal of Robotics Research 33~(2) (2014) 268--304.

\bibitem{luders_rrt}
B.~Luders, M.~Kothari, J.~How, Chance constrained rrt for probabilistic
  robustness to environmental uncertainty, in: AIAA guidance, navigation, and
  control conference, 2010, p. 8160.

\bibitem{luders_rrtstar}
B.~D. Luders, S.~Karaman, J.~P. How, Robust sampling-based motion planning with
  asymptotic optimality guarantees, in: AIAA Guidance, Navigation, and Control
  (GNC) Conference, 2013, p. 5097.

\bibitem{blackmore_cc_mp}
L.~Blackmore, M.~Ono, B.~C. Williams, Chance-constrained optimal path planning
  with obstacles, IEEE Transactions on Robotics 27~(6) (2011) 1080--1094.

\bibitem{liu_risk_aware_mp}
W.~Liu, M.~H. Ang, Incremental sampling-based algorithm for risk-aware planning
  under motion uncertainty, in: 2014 IEEE International Conference on Robotics
  and Automation (ICRA), IEEE, 2014, pp. 2051--2058.

\bibitem{zhu2019chance}
H.~Zhu, J.~Alonso-Mora, Chance-constrained collision avoidance for mavs in
  dynamic environments, IEEE Robotics and Automation Letters 4~(2) (2019)
  776--783.

\bibitem{rockafellar2007coherent}
R.~T. Rockafellar, Coherent approaches to risk in optimization under
  uncertainty, in: OR Tools and Applications: Glimpses of Future Technologies,
  Informs, 2007, pp. 38--61.

\bibitem{pavone_risk}
A.~Majumdar, M.~Pavone, How should a robot assess risk? towards an axiomatic
  theory of risk in robotics, in: Robotics Research, Springer, 2020, pp.
  75--84.

\bibitem{dr_goh}
J.~Goh, M.~Sim, Distributionally robust optimization and its tractable
  approximations, Operations research 58~(4-part-1) (2010) 902--917.

\bibitem{florenceIPC}
P.~Florence, J.~Carter, R.~Tedrake, Integrated perception and control at high
  speed: Evaluating collision avoidance maneuvers without maps, in: Workshop on
  the Algorithmic Foundations of Robotics (WAFR), 2016.

\bibitem{summers_iros_2018}
T.~Summers, Distributionally robust sampling-based motion planning under
  uncertainty, in: 2018 IEEE/RSJ International Conference on Intelligent Robots
  and Systems (IROS), IEEE, 2018, pp. 6518--6523.

\bibitem{davide_perception_plan}
G.~Costante, C.~Forster, J.~Delmerico, P.~Valigi, D.~Scaramuzza,
  Perception-aware path planning, arXiv preprint arXiv:1605.04151 (2016).

\bibitem{renganathan2020integrated}
V.~Renganathan, I.~Shames, T.~H. Summers, Towards integrated perception and
  motion planning with distributionally robust risk constraints,
  IFAC-PapersOnLine 53~(2) (2020) 15530--15536, 21th IFAC World Congress.
\newblock \href {https://doi.org/https://doi.org/10.1016/j.ifacol.2020.12.2396}
  {\path{doi:https://doi.org/10.1016/j.ifacol.2020.12.2396}}.

\bibitem{hakobyan2020wasserstein}
A.~Hakobyan, I.~Yang, Wasserstein distributionally robust motion control for
  collision avoidance using conditional value-at-risk, arXiv preprint
  arXiv:2001.04727 (2020).

\bibitem{safaoui2021risk}
S.~Safaoui, B.~J. Gravell, V.~Renganathan, T.~H. Summers, Risk-averse rrt*
  planning with nonlinear steering and tracking controllers for nonlinear
  robotic systems under uncertainty, arXiv preprint arXiv:2103.05572 (2021).

\bibitem{carla_sim}
A.~Dosovitskiy, G.~Ros, F.~Codevilla, A.~Lopez, V.~Koltun, Carla: An open urban
  driving simulator, in: Conference on robot learning, PMLR, 2017, pp. 1--16.

\bibitem{semantic_slam}
N.~S{\"u}nderhauf, T.~T. Pham, Y.~Latif, M.~Milford, I.~Reid, Meaningful maps
  with object-oriented semantic mapping, in: 2017 IEEE/RSJ International
  Conference on Intelligent Robots and Systems (IROS), IEEE, 2017, pp.
  5079--5085.

\bibitem{gu2005}
D.~Gu, H.~Hu, A stabilizing receding horizon regulator for nonholonomic mobile
  robots, IEEE Transactions on Robotics 21~(5) (2005) 1022--1028.

\bibitem{Allgower_NMPC}
R.~Findeisen, L.~Imsland, F.~Allgower, B.~A. Foss, State and output feedback
  nonlinear model predictive control: An overview, European Journal of Control
  9~(2) (2003) 190 -- 206.
\newblock \href {https://doi.org/https://doi.org/10.3166/ejc.9.190-206}
  {\path{doi:https://doi.org/10.3166/ejc.9.190-206}}.

\bibitem{ukf_correl_noises}
G.~Chang, Marginal unscented kalman filter for cross-correlated process and
  observation noise at the same epoch, IET Radar, Sonar \& Navigation 8~(1)
  (2014) 54--64.

\bibitem{wan2000unscented}
E.~A. Wan, R.~Van Der~Merwe, The {U}nscented {K}alman {F}ilter for nonlinear
  estimation, in: Proceedings of the IEEE 2000 Adaptive Systems for Signal
  Processing, Communications, and Control Symposium (Cat. No. 00EX373), IEEE,
  2000, pp. 153--158.

\bibitem{frazzoli_real_mp}
E.~Frazzoli, M.~A. Dahleh, E.~Feron, Real-time motion planning for agile
  autonomous vehicles, Journal of guidance, control \& dynamics 25~(1) (2002)
  116--129.

\bibitem{park_nonhol_dist}
J.~J. Park, B.~Kuipers, Feedback motion planning via non-holonomic rrt* for
  mobile robots, in: 2015 IEEE/RSJ International Conference on Intelligent
  Robots and Systems (IROS), IEEE, 2015, pp. 4035--4040.

\bibitem{revisiting_rrtstar}
K.~Solovey, L.~Janson, E.~Schmerling, E.~Frazzoli, M.~Pavone, Revisiting the
  asymptotic optimality of {RRT*}, in: 2020 IEEE International Conference on
  Robotics and Automation (ICRA), IEEE, 2020, pp. 2189--2195.

\bibitem{drcclp}
G.~C. Calafiore, L.~{El Ghaoui}, On distributionally robust chance-constrained
  linear programs, Jour. of Optimization Theory and Applications 130~(1) (Dec.
  2006).

\bibitem{casadi}
J.~A.~E. Andersson, J.~Gillis, G.~Horn, J.~B. Rawlings, M.~Diehl, {CasADi} --
  {A} software framework for nonlinear optimization and optimal control,
  Mathematical Programming Computation 11~(1) (2019) 1--36.
\newblock \href {https://doi.org/10.1007/s12532-018-0139-4}
  {\path{doi:10.1007/s12532-018-0139-4}}.

\bibitem{kongbicycle}
J.~Kong, M.~Pfeiffer, G.~Schildbach, F.~Borrelli, Kinematic and dynamic vehicle
  models for autonomous driving control design, in: 2015 IEEE Intelligent
  Vehicles Symposium (IV), IEEE, 2015, pp. 1094--1099.

\end{thebibliography}

\end{document}